%% file: main.tex
\documentclass[sigconf,balance=false]{acmart}

\iftrue
\setcopyright{none}
\renewcommand\footnotetextcopyrightpermission[1]{} 
\fi

\def\Comments{0} 
\input{macros}

\begin{document}


\title[Summary Reports Optimization in the Privacy Sandbox Attribution Reporting API]{Summary Reports Optimization\\ in the Privacy Sandbox Attribution Reporting API}


\author{
\begin{tabular}{cccc}
Hidayet Aksu& Badih Ghazi& Pritish Kamath& Ravi Kumar
\end{tabular}
\begin{tabular}{ccc}
Pasin Manurangsi& Adam Sealfon& Avinash V Varadarajan
\end{tabular}
}
\affiliation{
\institution{Google}
\country{}}
\email{ {aksu, pritishk, pasin, adamsealfon, avaradar}@google.com, {badihghazi,ravi.k53}@gmail.com }




\renewcommand{\shortauthors}{}


\begin{abstract}
The Privacy Sandbox Attribution Reporting API has been recently deployed by Google Chrome to support the basic advertising functionality of attribution reporting (aka conversion measurement) after deprecation of third-party cookies. The API implements a collection of privacy-enhancing guardrails including contribution bounding and noise injection. It also offers flexibility for the analyst to allocate the contribution budget.

In this work, we present methods for optimizing the allocation of the contribution budget for summary reports from the Attribution Reporting API. We evaluate them on real-world datasets as well as on a synthetic data model that we find to accurately capture real-world conversion data. Our results demonstrate that optimizing the parameters that can be set by the analyst can significantly improve the utility achieved by querying the API while satisfying the same privacy bounds.

\end{abstract}


\maketitle
\pagestyle{plain}

\section{Introduction}
In recent years, growing concerns around user privacy have led to new efforts by web browsers and mobile platforms to limit pervasive tracking of users by websites, apps, and ad technology (aka, \emph{ad-tech}) providers. In particular, this led to the decision by several browsers and platforms, including Safari \cite{safari}, Mozilla Firefox \cite{mozilla}, and Google Chrome \cite{chromium}, to deprecate third-party cookies. However, third-party cookies had been widely used to support some of the most critical functionalities powering digital advertising, notably ad conversion measurement (aka attribution reporting), where an ad-tech seeks to determine the volume of conversions attributed to ads shown on different publishers and as part of different campaigns.  Example conversion measurement queries include the number of conversions attributed to ad impressions shown on a given publisher, or the total conversion value for sales occurring during a weekend and attributed to a particular ad campaign. This made several platforms and browsers provide privacy-preserving APIs that can support ad conversion measurement functionalities after the deprecation of third-party cookies, including Private Click Measurement (PCM) on Safari \cite{pcm-safari}, SKAdNetwork on iOS \cite{SKAdNetwork}, the Interoperable Private Attribution (IPA) developed by Mozilla and Meta \cite{ipa-blog},  Masked LARK from Microsoft \cite{pfeiffer2021masked, MaskedLARk-github}, and Privacy Sandbox Attribution Reporting API (ARA) from Google~\cite{chrome-attribution-reporting, aggregate-api-android}.

The Attribution Reporting API, available on both the Chrome browser \cite{chrome-attribution-reporting} and the Android operating system \cite{aggregate-api-android}, offers \emph{summary reports} that could be used to estimate counts and values of conversions attributed to ad campaigns (and broken down by other impression and conversion features). The privacy guardrails in the API \cite{privacy-guardrails-aggregate-api} include contribution bounding as well as discrete Laplace noise injection, which can be used to provide a differential privacy (DP) \cite{DworkMNS06} guarantee on the output summary reports. More precisely, for each impression, the API enforces a fixed bound on the contributions of all conversions attributed to it. Moreover, each of these contributions is required to be discrete. The contributions from different impressions (from possibly many users) are aggregated. Discrete Laplace noise is then added to the vector of contributions, and the result is the summary report. 

While ARA provides a formal differential privacy guarantee, the noise addition and contribution bounding procedure represents a paradigm shift in conversion measurement compared to third-party cookies; straightforward use of ARA might result in large amount of noise. This degradation in accuracy can in turn impact the downstream business decisions made based on the measurements.

This brings us to the main question of the paper:
\emph{How can an ad-tech obtain desired measurements that are as accurate as possible via ARA?} The flexibility of ARA allows the ad-tech to choose their own encoding of the attributed information. (See \Cref{sec:ara_summary_reports_constraints} for formal descriptions.) By adjusting such an encoding, the ad-tech can (implicitly) decide on several parameters, such as how the contribution budget is allocated across different conversions that are attributed to the same impression. These choices on the ad-tech part can significantly impact the utility of the summary reports for any fixed privacy bar. The optimization of these summary reports so as to maximize utility for a given level of (differential) privacy is the focus of this work.


{\fontsize{5}{7}\selectfont
\newcommand{\impCol}[1]{\textcolor{black!30!Gred}{#1}}
\newcommand{\campaignCol}[1]{\textcolor{black!30!Gred}{#1}}
\newcommand{\locationCol}[1]{\textcolor{black!30!Gred}{#1}}
\newcommand{\convCol}[1]{\textcolor{black!30!Ggreen}{#1}}
\newcommand{\histCol}[1]{\textcolor{black!30!Gyellow}{#1}}
\begin{figure}[t]
\centering
\begin{tikzpicture}
\tikzset{
    box/.style = {rectangle, rounded corners=6pt, line width=0.7pt, draw, inner sep=5pt}
}
\node[box, text width=3.4cm, text depth=2.8cm, below right, draw=black!30!Gred] (imp) at (0, 0) {
    {\fontsize{5.5}{7}\selectfont%
    \impCol{\bf Impression}\\
    \impCol{\bf (Campaign = Thanksgiving)}\\[2.5mm]
    }
    
    {\bf aggregation\_keys:}\\
    ``Items, Campaign'': ``\impCol{\texttt{Items, Thanksgiving}}''\\
    ``Value, Campaign'': ``\impCol{\texttt{Value, Thanksgiving}}''\\
    ``$\bot$, Campaign'': ``\impCol{\texttt{$\bot$, Thanksgiving}}''\\
};
\node[text width=8mm, above left] at ($(imp.south east)+(-0.1,0.1)$){\bf
    Ad-tech\\[-0.5mm]
    Publisher
};

\node[box, text width=3.4cm, text depth=2.8cm, below right, draw=black!30!Ggreen] (conv) at ($(imp.north east) + (0.1,0)$) {
    {\fontsize{5.5}{7}\selectfont%
    \convCol{\bf Conversion}\\
    \convCol{\bf (\#items = 3, value = \$21)}\\[2.5mm]
    }
    
    {\bf aggregatable\_trigger\_data:}\\
    key\_piece:  ``'', source\_key: ``Items, Campaign''\\
    key\_piece: ``'', source\_key: ``Value, Campaign''\\
    key\_piece: ``'', source\_key: ``$\bot$, Campaign''\\[2.5mm]
    
    {\bf aggregatable\_values:}\\
    ``Items, Campaign'': \convCol{16384}\\
    ``Value, Campaign'': \convCol{11469}\\
    ``$\bot$, Campaign'': \convCol{4915}
};
\node[text width=8mm, above left] at ($(conv.south east)+(-0.1,0.1)$){\bf
    Ad-tech\\[-0.5mm]
    Advertiser
};

\node[box, text width=3.4cm, text depth=1.3cm, below right, draw=black!30!Gyellow] (hist) at (2.4, -4) {
    {\fontsize{5.5}{7}\selectfont%
    \histCol{\bf Histogram Contributions}\\[2.5mm]
    }
    
    key: ``\impCol{\texttt{Items, Thanksgiving}}'', value: \convCol{16384}\\
    key: ``\impCol{\texttt{Value, Thanksgiving}}'', value: \convCol{11469}\\
    key: ``\impCol{\texttt{$\bot$, Thanksgiving}}'', value: \convCol{4915}\\
    };
    \node[text width=6mm, above left] at ($(hist.south east)+(-0.1,0.1)$){\bf
    Browser
};

\path[-{Latex[width=1.5mm,length=1.5mm]}, line width=0.7pt]
(imp) edge (hist)
(conv) edge (hist);
\end{tikzpicture}
\caption{Histogram Contributions Generation.}
\label{fig:hist_contrib}
\end{figure}
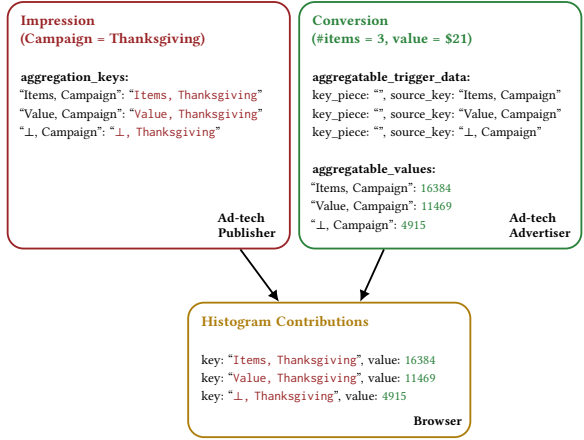
}

\newcommand{\GiftShop}{{\em Du \& Penc}\xspace}

\begin{table*}[htb]
\centering\small
\begin{tabular}{c|c|c|c|c|c|}
\cline{2-6}
& \multicolumn{3}{|c|}{\boldmath\bf Impression features $x$} & \multicolumn{2}{c|}{\boldmath\bf Conversion features $y$}\\
\cline{2-6}
& Impression ID & Campaign & City & \#items & value (\$) \\
\cline{2-6}
$z_1 \rightarrow$ & {\tt 123} & {\tt Thanksgiving} & New York & 3 & 21\\
$z_2 \rightarrow$ & {\tt 123} & {\tt Thanksgiving} & New York & 1 & 5\\
$z_3 \rightarrow$ & {\tt 456} & {\tt Thanksgiving} & Boston & 1 & 99\\
$z_4 \rightarrow$ & {\tt 123} & {\tt Thanksgiving} & New York & 2 & 23\\
$z_5 \rightarrow$ & {\tt 101} & {\tt Christmas} & Boston & 2 & 50\\
$z_6 \rightarrow$ & {\tt 789} & {\tt Christmas} & New York & 3 & 15\\
$z_7 \rightarrow$ & {\tt 101} & {\tt Christmas} & Boston & 1 & 5\\
\ldots & \ldots & \ldots & \ldots & \ldots & \ldots \\
\cline{2-6}
\end{tabular}
\caption{Running example showing the impression and conversion logs for \GiftShop, an online gift shop.}
\label{tab:dataset_example}
\end{table*}

\subsection{Summary of Contributions}
We make the following contributions to the problem of optimizing summary reports from the Attribution Reporting API (ARA):
\begin{itemize}[leftmargin=*]
\item We formally define the problem of optimizing the utility of conversion aggregates based on summary reports from the ARA. In the process, we flesh out subtle but important details regarding how the constraints in the API (e.g., contribution bounding, discretization requirements, encoding of multiple conversions attributed to the same impression, noise injection) can shape the optimization problem. While our framework allows for a variety of error measures, we discuss several qualitative advantages of using a thresholded version of the root mean square relative error error in ad conversion measurement, compared to other error measures, including the more standard additive and relative (aka multiplicative) error measures.
\item Given access to historical data that has not been contribution-bounded and is noise-free, we provide an optimization problem that yields the optimal choice of parameters to use the ARA on future data. Such historical data could be available to ad-techs who have thus far relied on third-party cookies for ad conversion measurement (prior to the deprecation of third-party cookies).
\item We evaluate our algorithm on real-world conversion data, demonstrating that it significantly improves utility compared to baseline non-optimized summary reports.  We also evaluate our algorithm on synthetic datasets (for attributed conversion counts and values), which are sampled using generative models that we fit to real conversion data. These data generation models might be of independent interest for future research on (privacy-preserving) ad conversion measurement.
\item As our algorithm uses past (historical) data in order to set the parameters used to measure conversions on future data, we complement our empirical findings by proving generalization bounds showing that parameters optimized using historical bounds yield good results on similar future data.
\end{itemize}


We remark that, due to the rather specific nature of the Attribution Reporting API (and, to a lesser extent, differentially private conversion measurements), we are not aware of any previous work that studies the same setting as ours. Nonetheless, we discuss some related work in \Cref{sec:related_work}.

\subsection{Overview of the Rest of the Paper}
In \Cref{sec:prelim}, we start with some basic definitions related to ad conversion measurement, and to differential privacy. In \Cref{sec:additional-background}, we provide further background on the problem setup; ARA is formalized in more detail in \Cref{sec:ara_summary_reports_constraints}, the problem of estimating conversion aggregates (the focus of our work) is defined in \Cref{sec:problem}, and we briefly discuss the error metric in \Cref{subsec:error_metric}. 
%
In \Cref{sec:capping}, we present our optimization algorithm for contribution budgeting. We describe our experimental evaluation and findings in \Cref{sec:experiments}. In \Cref{sec:generalization}, we prove generalization bounds which explain why our algorithm does not overfit to the historical data. We discuss some related work in \Cref{sec:related_work}, and multiple interesting research directions in \Cref{sec:conc_future_directions}.

\section{Preliminaries}\label{sec:prelim}

We now define the main terminology used in this paper. Let $\cX$ be the set of {\em impression features}, including a unique impression identifier for each impression, let $\cY$ be the set of {\em conversion features}, and let $\cZ = \cX \times \cY$.
We denote a dataset as $D = (z_1, \ldots, z_n) \in \cZ^*$ where each {\em record} $z_i$ is of the form $(x_i, y_i)$,
consisting of {\em impression features} $x_i \in \cX$ and {\em conversion features} $y_i \in \cY$. The impression features are assumed to be {\em known} to the ad-tech, and can correspond to an a priori unbounded number of conversions, which are assumed to be {\em unknown} to the ad-tech. We say that two datasets $D, D'$ are {\em adjacent}, denoted as $D \sim D'$ if we can get one dataset from the other by removing all records corresponding to a single impression.\pritishinfo{Under special conditions, we can actually handle substitution adjacency, but we can ignore those for simplicity.}

We use the following running example to illustrate our notation. \GiftShop
is a gift shop, which uses digital advertising to reach its customers. Their holiday sales are captured in the dataset in \Cref{tab:dataset_example}, where each record contains impression features of (i) a {unique}\hidayet{In the table Impression IDs are not unique, can we drop word `unique` here? (may confuse reader)} impression ID, (ii) the campaign, and (iii) the city in which the ad was shown, as well as the conversion features of the (i) number of items bought and (ii) total dollar value of items bought as part of the conversion.

\subsection{Differential Privacy}\label{subsec:dp_basics}

\begin{definition}[DP~\cite{DworkMNS06}]\label{def:differential_privacy}
For $\eps \geq 0$, a randomized algorithm $\cA$ is \emph{$\eps$-DP} if for all adjacent datasets $D \sim D'$, and for every possible output $o$, it holds that $\Pr[\cA(D) = o] \leq e^\eps \cdot \Pr[\cA(D') = o]$.
\end{definition}

For an extensive overview of DP, we refer the reader to the monograph \cite{dwork2014algorithmic}.
A commonly used method in DP is the discrete Laplace mechanism. To define it, we recall the notion of $\ell_1$-sensitivity, where for any vector $v \in \mathbb{R}^d$, we denote its \emph{$\ell_1$-norm} as $\|v\|_1 := \sum_{i=1}^d |v_i|$.
\begin{definition}[$\ell_1$-sensitivity]
Let $\cZ$ be any set, and $f: \cZ^n \to \R^d$ be a $d$-dimensional function. Its \emph{$\ell_1$-sensitivity} is defined as $\Delta_1 f := \max_{D \sim D'} \|f(D) - f(D')\|_1$.
\end{definition}

\begin{definition}[Discrete Laplace Mechanism]\label{def:disc_lap_mech}
The \emph{discrete Laplace distribution} centered at $0$ and with parameter $a > 0$, denoted by $\DLap(a)$, is the distribution whose probability mass function at integer $k$ is $\frac{e^{a} - 1}{e^{a} + 1} \cdot e^{- a|k|}$.
The \emph{$d$-dimensional discrete Laplace mechanism} with parameter $a$ applied to a function $f: \cZ^n \to \Z^d$, on input a dataset $D \in \cZ^n$, returns $f(D) + \xi$ where $\xi$ is a $d$-dimensional noise random variable whose coordinates are sampled i.i.d. from $\DLap(a)$ (abbreviated as $\xi \sim \DLap(a)^{\otimes d}$).
\end{definition}

\begin{lemma}\label{lem:dp_discrete_laplace}
For all $\eps > 0$, the $d$-dimensional discrete Laplace mechanism with parameter $a \le \eps / \Delta_1 f$ is $\eps$-DP.
\end{lemma}

\begin{lemma}\label{lem:var_discrete_laplace}
For all $a > 0$, it holds that\pritish{Is there a reference? I could not find a reference online for the variance, so used Mathematica to compute it.}
\[
\Ex[\DLap(a)] ~=~ 0 \ \ \text{and} \ \  \Var(\DLap(a)) ~=~ \frac{2 e^a}{(e^a - 1)^2}\,.
\]
\end{lemma}

The following is a well-known property of DP.
\begin{lemma}[Post-processing]\label{lem:dp_post_processing}
If $\cA$ is $\eps$-DP, then for any (randomized) algorithm $\cA'$, it holds that $\cA'(\cA(\cdot))$ is also $\eps$-DP.
\end{lemma}


\section{Background and Setup}
\label{sec:additional-background}

\subsection{ARA Summary Reports}\label{sec:ara_summary_reports_constraints}

We now describe {\em Summary Reports}, which is a pre-defined framework for ad measurement fixed in the ARA.\pritishinfo{I added this line to emphasize that the Summary Report framework is pre-defined, not our creation. Please modify as appropriate.} At a high-level, ARA allows the ad-tech to specify the encoding algorithm from each record to a vector. A contribution bounding procedure is then applied to ensure that the total contribution corresponding to each impression is bounded. Once this is done, they are sent to the aggregation service who sums these vectors and adds a discrete Laplace noise to the result. This (noisy) summary report is then returned to the ad-tech.

More formally, a mechanism $M$ using ARA summary reports operates as follows.
To begin with, the ARA has a parameter called \emph{contribution budget}, which at this time of writing, is fixed to $\contribbudget := 2^{16} = 65,536$ (see \cite{contribution-budget-ara}).
To use the API, the ad-tech needs to specify the following information beforehand:
\begin{itemize}[leftmargin=4mm,topsep=3pt]
\item a set $K = \{k_1, \ldots, k_T\}$ of {\em aggregation keys}, and
\item an encoding algorithm $\cA$ that maps any record $z$ to a {\em histogram contribution}
$w^z \in \Z_{\ge 0}^K$, where $w^z_k$ is called the {\em aggregatable value} cooresponding to the {\em aggregation key} $k$.
\end{itemize}
Let $\cZ^x := (z_{i_1}, \ldots, z_{i_c})$ be the sequence of records corresponding to the same impression $x$, i.e., $z_{i_j} = (x, y_{i_j})$, and let $\cW^x$ be the sequence of corresponding histogram contributions $(w^{z_{i_1}}, \ldots, w^{z_{i_c}})$.
The histogram contributions of impression $x$ are then filtered\footnote{\Cref{alg:contribution-capping} as written requires all the inputs to be provided at once. But in practice, it runs in an online manner, namely, the $w_i$s arrive sequentially and the decision of whether to include $w_i$ or not, is made without knowing the future $w_{i'}$s, and the algorithm only needs to remember a single $16$-bit value $b$ for each impression $x$.} by $\cC$ (\Cref{alg:contribution-capping}), such that the $\ell_1$-norm of the sum of the returned vectors, called {\em aggregatable reports}, is at most $\contribbudget$. For ease of notation, we use $\ocW_x$ to denote the sequence of aggregatable reports for impression $x$, and let $\ow^x := \sum_{w \in \ocW^x} w$. Similarly, we use $\ocZ^x$ to denote the sequence of records corresponding to aggregatable reports for impression $x$.

The aggregatable reports are then passed on by $\cC$ to the ARA {\em aggregation service} $\cS$ that adds all the aggregatable reports and adds discrete Laplace noise, as given in \Cref{alg:summary-report}, and returns a {\em summary report} $W \in \Z^K$ back to the ad-tech.

\begin{figure*}
\centering
\begin{tikzpicture}
\def\xZ{-8.9}
\def\xA{-7.6}
\def\xC{-5}
\def\xS{-2.4}
\def\xR{0.1}
\def\xU{2.1}

\node[rectangle, rounded corners=12pt, draw=none, fill=Ggreen!20, line width=0.8pt, text width=4.9cm, text depth=6.4cm, align=center, below] at (0.5*\xS+0.5*\xC-0.1, 3.6) {
	\mbox{}\\[1mm]\color{black!30!Ggreen}\fontsize{10pt}{10}\selectfont\bf
	Attribution Reporting API
};

\tikzset{
    env/.style = {rectangle, rounded corners=3pt, draw=none, fill=black!20, fill opacity=0.5, align=center, below},
    client/.style = {env, text width=7.2cm}
}
\node[client, text depth=1.8cm] (CL1) at (0.5*\xC+0.5*\xZ-1.2, 2.7) {};
\node[client, text depth=0.6cm] (CL2) at (0.5*\xC+0.5*\xZ-1.2, 0.4) {};
\node[client, text depth=1.15cm] (CL3) at (0.5*\xC+0.5*\xZ-1.2, -0.725) {};
\node[left] at ($(CL1.west)+(0.8,0)$) {\includegraphics[height=5mm]{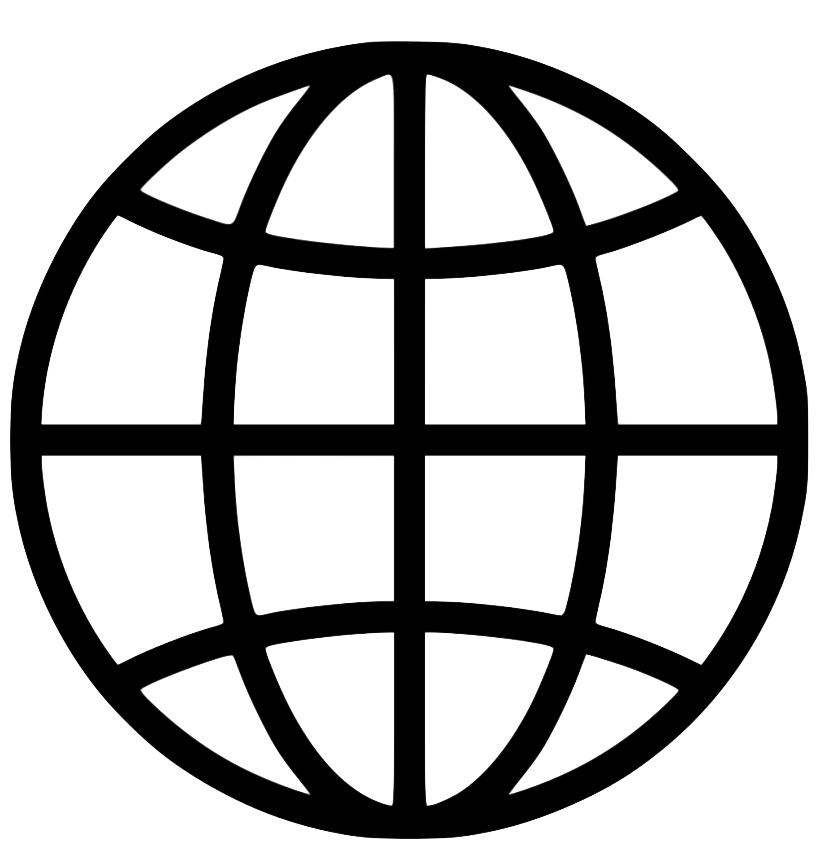}};
\node[left] at ($(CL2.west)+(0.8,0)$) {\includegraphics[height=5mm]{figs/browser_logo.png}};
\node[left] at ($(CL3.west)+(0.8,0)$) {\includegraphics[height=5mm]{figs/browser_logo.png}};

\node[env, text width=1.7cm, text depth=2.5cm] (CL1) at (\xS, 2) {\mbox{}\\[-1mm]
    \fontsize{7.5pt}{10}\selectfont\bf
    Trusted\\
    Execution\\
    Environment
};

\node[env, text width=3.2cm, text depth=1.8cm] (CL1) at (\xR+0.9, 1.3) {\mbox{}\\[-1mm]
    \fontsize{7.5pt}{10}\selectfont\bf
    Ad-tech
};

\tikzset{
    algline/.style = {dotted, line width=0.8pt}
}
{
\fontsize{7pt}{10}\selectfont
\node[text width=4cm, align=center] at (\xA,-3.2) {\HistogramContribution\\ (\Cref{alg:contribution-vector-Linf})} edge[algline, Gred] (\xA, 2.9);
\node[text width=4cm, align=center] at (\xC,-2.5) {\ContributionCapping\\ (\Cref{alg:contribution-capping})} edge[algline, black!30!Gyellow] (\xC, 2.3);
\node[text width=4cm, align=center] at (\xS,-1.4) {\SummaryReport\\ (\Cref{alg:summary-report})} edge[algline, Gblue] (\xS, 0.6);
\node[text width=4cm, align=center] at (\xR,-1.4) {\ReconstructValues\\ (\Cref{alg:reconstruct})} edge[algline, myPurple] (\xR, 0.6);
}

\tikzset{
    box/.style = {rectangle, rounded corners=4pt, draw=\BoxCol, fill=\BoxCol!30, line width=0.8pt, inner sep=4pt},
    arrow/.style = {-{Stealth}, line width=0.7pt}
}

{\fontsize{18pt}{0}\selectfont
\node (U) at (\xU, 0) {$U$};

\def\BoxCol{myPurple}
\def\xgap{\xR-\xU}
\node[box] (R) at ($(U)+(\xgap,0)$) {$\cR$} edge[arrow] (U);

\def\BoxCol{Gblue}
\def\xgap{\xS-\xR}
\node[box] (S) at ($(R)+(\xgap,0)$) {$\cS$} edge[arrow] (R);
}

{\fontsize{11pt}{0}\selectfont
\def\BoxCol{Gyellow}
\def\xgap{\xC-\xS}
\node[box] (C1) at ($(S)+(\xgap,1.7)$) {$\cC$} edge[arrow] (S);
\node[box] (C2) at ($(S)+(\xgap,0)$) {$\cC$} edge[arrow] (S);
\node[box] (C3) at ($(S)+(\xgap,-1.4)$) {$\cC$} edge[arrow] (S);
}

\def\BoxCol{Gred}
\def\xgap{\xA-\xC}
\node[box] (A11) at ($(C1)+(\xgap,0.6)$) {$\cA$} edge[arrow] (C1);
\node[box] (A12) at ($(C1)+(\xgap,0)$) {$\cA$} edge[arrow] (C1);
\node[box] (A13) at ($(C1)+(\xgap,-0.6)$) {$\cA$} edge[arrow] (C1);
\node[box] (A21) at ($(C2)+(\xgap,0)$) {$\cA$} edge[arrow] (C2);
\node[box] (A31) at ($(C3)+(\xgap,0.3)$) {$\cA$} edge[arrow] (C3);
\node[box] (A32) at ($(C3)+(\xgap,-0.3)$) {$\cA$} edge[arrow] (C3);

\def\xgap{\xZ-\xA}
\node[left] (z11) at ($(A11)+(\xgap,0)$) {$(x_1, y_{1,1}) = z_{1,1}$} edge[arrow] (A11);
\node[left] (z12) at ($(A12)+(\xgap,0)$) {$(x_1, y_{1,2}) = z_{1,2}$} edge[arrow] (A12);
\node[left] (z13) at ($(A13)+(\xgap,0)$) {$(x_1, y_{1,3}) = z_{1,3}$} edge[arrow] (A13);
\node[left] (z21) at ($(A21)+(\xgap,0)$) {$(x_2, y_{2,1}) = z_{2,1}$} edge[arrow] (A21);
\node[left] (z31) at ($(A31)+(\xgap,0)$) {$(x_3, y_{3,1}) = z_{3,1}$} edge[arrow] (A31);
\node[left] (z32) at ($(A32)+(\xgap,0)$) {$(x_3, y_{3,2}) = z_{3,2}$} edge[arrow] (A32);
\end{tikzpicture}
\caption{\boldmath Illustrative usage of ARA summary reports. Algorithms $\cC$ and $\cS$ are fixed in the API. The ad-tech designs $\cA$ and $\cR$.}
\label{fig:ara_algorithms}
\end{figure*}

Note that algorithm $\cA$ and $\cC$ are executed on the browser/device on which the impression occurred, and only the aggregatable reports are passed on to $\cS$ to produce the summary report, which is executed in a trusted execution environment~\cite{aggregation-service-tee}, as illustrated in \Cref{fig:ara_algorithms}. Finally, the ad-tech can post-process the summary report $W$ using any algorithm $\cR$ to obtain the final estimate $U \in \R^{d \times m}$.

\begin{algorithm}[t]
\caption{\ContributionCapping $\cC$.}
\label{alg:contribution-capping}
\begin{algorithmic}
\STATE {\bf Params:} Contribution budget $\contribbudget := 2^{16} = 65,536$
\STATE {\bf Input:} Sequence of {\em histogram contributions} $w_1, \ldots, w_c \in \Z_{\ge 0}^K$ corresponding to a single impression
\STATE {\bf Output:} Sequence of {\em aggregatable reports} $w_1, \ldots, w_{\hat{c}} \in \Z_{\ge 0}^K$ such that $\|\sum_{j=1}^{\hat{c}} w_j\|_1 \le \contribbudget$
\STATE 
\STATE $S \gets \emptyset \text{ and } b \gets 0$
\FOR{$i \in \{1, \ldots, c\}$}
    \IF{$b + \|w_i\|_1 \le \contribbudget$}
        \STATE $S \gets S \cup \{i\} \text{ and } b \gets b + \|w_i\|_1$
    \ENDIF
\ENDFOR
\RETURN $(w_i)_{i \in S}$
\end{algorithmic}
\end{algorithm}

\begin{algorithm}[t]
\caption{\SummaryReport $\cS$.}
\label{alg:summary-report}
\begin{algorithmic}
\STATE {\bf Input:} Aggregatable reports $w_1, \ldots, w_{r} \in \Z_{\ge 0}^K$
\STATE {\bf Output:} Summary report $W \in \Z^{K}$
\STATE 
\RETURN $W \gets \sum_{j} w_j + \DLap(\eps / \contribbudget)^{\otimes K}$
\end{algorithmic}
\end{algorithm}

An important point to note is that algorithms $\cC$ and $\cS$ are fixed in ARA, and the only parts that the Ad-tech can control are algorithms $\cA$ and $\cR$. The design of ARA summary reports ensures that, no matter what algorithms $\cA$ and $\cR$ are provided, it is ensured that the final result received by the Ad-tech satisfies $\eps$-DP. We include a short proof of this statement below.

\begin{theorem}\label{thm:ara-dp}
For all $\cA$ and $\cR$, the final output received by the Ad-tech satisfies $\eps$-DP.\pritish{Might want to rephrase this theorem statement.}
\end{theorem}
\begin{proof}
Since each impression $x$ contributes $\ow^x$ to $W$ with $\|\ow^x\|_1 \le \contribbudget$, it follows that the summary report $W$ is guaranteed to be $\eps$-DP (\Cref{lem:dp_discrete_laplace}) under adding or removing all conversions attributed to a single impression.\footnote{Note that this is a simpler setting compared to adding or removing an impression, which may create a ``cascading effect'' that leads to modifications in the conversions attributed to multiple impressions at once. However, since the current version of the ARA does not touch upon this issue, we will do the same in this paper.} Finally, algorithm $\cR$ only post-processes the summary report $W$ to obtain the final estimate $U \in \R^{d \times m}$, and hence by \Cref{lem:dp_post_processing} this preserves $\eps$-DP.
\end{proof}

Recall that, the goal of our paper is to design a mechanism in this framework that minimizes the error (in particular $\RMSRE_{\btau}$). In order to design a mechanism, we need to specify the following: (i) the set $K$ of aggregation keys and (ii) the histogram contributions $w^z$. As we will demonstrate through the rest of this paper, careful selection of these can lead to significant utility improvement.


\subsection{Estimating Conversion Aggregates}\label{sec:problem}

As mentioned earlier, we are interested in computing aggregate statistics on attributed conversion where they can be ``sliced'' based on certain attributes. This problem is formalized below.


Consider a fixed partition of $\cZ$ given as $\cZ_1 \sqcup \cdots \sqcup \cZ_m$, where the $\cZ_j$'s are pairwise disjoint (where $\sqcup$ denotes a disjoint union).
This partition naturally induces a partition of $D$ given as $D_1 \sqcup \cdots \sqcup D_m$ where $D_j = D \cap \cZ_j$; we refer to each $D_j$ as a {\em slice} of $D$. We use $X := \{ x : \exists\, y \in \cY \text{ s.t. } (x,y) \in D \}$ and $X_j := \{ x : \exists\, y \in \cY \text{ s.t. } (x,y) \in D_j \}$, whenever $D$ is clear from context.

A {\em query} is defined by a function $q : \cZ \to \R$.
The {\em aggregate value} associated with a query $q$ on dataset $D$ is $V_D(q) \in \R^m$, given as $V_D(q)_{j} := \sum_{z \in D_j} q(z)$, namely one aggregate for each slice of $D$.
Given queries $q_1, \ldots, q_d$ of interest, the goal is to construct an $\eps$-DP mechanism that estimates the corresponding aggregate values $V_D(q_1), \ldots, V_D(q_d)$ as ``accurately'' as possible.

In the \GiftShop example, one could consider a partition of the records, e.g., by {\em Campaign} or by {\em City} or by the pair {\em (Campaign, City)}. We could consider the following two queries: $q_1(z)$, which returns \#items and $q_2(z)$, which returns the dollar value in any record $z$. In addition, we always consider another query of interest, given as $q_0(z) = 1$ for all $z \in \cZ$; hence, $V_D(q_0)_j$ is precisely the number of records in the $j$th slice.

\newcommand{\RMSE}{\mathsf{RMSE}}

\subsection{Error Metrics for Experiment Evaluation}\label{subsec:error_metric}

Aggregate APIs can be used to generate privacy preserving reports on attributed conversions. However, since the reports are noisy, API users should evaluate the impact of noise in reports carefully to ensure that they are useful. A variety of utility metrics can be used to evaluate the impact of noise. We provide a list of all metrics that we have considered in \Cref{table:metrics_list} (\Cref{sec:error-metrics}). There are several desirable criteria for metrics. As listed in Table~\ref{table:metrics_features} (\Cref{sec:error-metrics}), we find out that only $\RMSRE_{\btau}$ metric (defined below) satisfies all desired properties. Therefore, $\RMSRE_{\btau}$ is used for evaluation in our experiments.

\begin{definition}[\cite{noise-lab}] \label{def:rmse-tau}
For a dataset $D \in \cZ^*$, a query $q : \cZ \to \R$, and a random vector $u \in \R^{m}$, the {\em root mean squared relative error} with parameter $\tau \in \R_{> 0}$ is defined as
\begin{align*}
\RMSRE_\tau(u, q; D) &\displaystyle~:=~ \sqrt{\frac{1}{m} \sum_{j \in [m]} \Ex \prn{\frac{u_{j} - V_D(q)_{j}}{\max\{\tau, V_D(q)_{j}\}}}^2},
\end{align*}
where the expectation is over the randomness of $u$.
Similarly, for parameters $\btau = (\tau_0, \ldots, \tau_d)$, queries $(q_0, q_1, \ldots, q_d)$, and a randomized report $U \in \R^{(d+1) \times m}$, we define
\begin{align*}
\RMSRE_\btau(U, (q_\ell)_{\ell=0}^d; D) &\displaystyle~:=~ \sqrt{\frac{1}{d+1} \sum_{\ell=0}^d \RMSRE_{\tau_\ell}(U_{\ell}, q_\ell; D)^2},
\end{align*}
where $U_{\ell}$ is the $\ell$th row of $U$.
\end{definition}


\section{Contribution Budgeting Algorithm} \label{sec:capping}
In this section, we present our algorithm for contribution budgeting. Suppose we have $d$ queries $q_1, \ldots, q_d : \cZ \to \R_{\ge 0}$ for which an ad-tech desires to estimate the aggregate values. We now describe our approach for defining the aggregation keys $K$, the encoding algorithm $\cA$ mapping records to histogram contributions, and the method $\cR$ for reconstructing the values.

\subsection{Aggregation Keys \& Histogram Contributions}\label{subsec:histogram_contributions}
We take the set of aggregation keys to be $K = ([d] \cup \{\bot\}) \times [m]$, i.e., we define $d+1$ aggregation keys corresponding to each slice $j \in [m]$.
We choose a {\em count limit} $C$, and additionally, corresponding to each query $q_\ell$, we choose a {\em clipping threshold} $C_\ell$\pritish{Note that due to stateless nature of the aggregate API, this is a {\em per conversion bounding} of value, and not a bound across all conversions.} and a {\em contribution budget fraction} $\alpha_\ell$, subject to $\sum_{\ell=1}^d \alpha_\ell = 1$. We discuss how to choose the parameters $C$ and $(C_\ell, \alpha_\ell)_{\ell \in [d]}$ shortly.

For any record $z$, suppose $v_1, \ldots, v_d$ are given as $v_\ell = q_\ell(z)$. We first clip each $v_\ell$ to be at most $C_\ell$, namely, let $v_\ell' = \clip_{C_\ell}(v_\ell) := \min(v_\ell, C_\ell)$. Next, we scale $v_\ell'$ to lie in $[0, 1]$, by dividing by $C_\ell$, namely, let $v_\ell'' = v_\ell' / C_\ell$. We will choose a histogram contribution $w^z$ that uses exactly $\floor{\contribbudget / C}$ of the total contribution budget, so that at most $C$ conversions can be accounted for per impression. To this end, we rescale $v_\ell''$ by $\floor{\alpha_\ell \contribbudget / C}$ so that the sum of contributions is at most $\contribbudget / C$, and moreover, we assign the remaining contribution mass to $\bot$. Finally, we apply a randomized rounding on the real values to make them integer-valued, as required by the API.

These steps are formalized in \Cref{alg:contribution-vector-Linf}, where $\CRR$ is a randomized method that {\em clips} and performs {\em randomized rounding}, defined as follows:
\begin{align}
\CRR(v; C, C_\ell, \alpha_\ell) &\textstyle~:=~ \RR\prn{\floor{\frac{\alpha_\ell \contribbudget}{C}} \cdot \frac{\clip_{C_\ell}(v)}{C_\ell}}, \\
\text{where }\quad
\RR(\omega) &\textstyle~:=~\begin{cases}
    \ceil{\omega} & \text{w.p. } \omega - \floor{\omega}\\
    \floor{\omega} & \text{w.p. } 1 - \omega + \floor{\omega}. \\
\end{cases}
\end{align}
We use $\RR$ instead of deterministic rounding because $\Ex[\RR(\omega)] = \omega$, which allows recovery of unbiased estimates from summary reports.

\begin{algorithm}[t]
\caption{\HistogramContribution $\cA$ ($\ell_\infty$ version).}
\label{alg:contribution-vector-Linf}
\begin{algorithmic}
\STATE {\bf Params:} $\triangleright$ Queries $q_1, \ldots, q_d : \cZ \to \R_{\ge 0}$
\STATE \phantom{\bf Params:} $\triangleright$ Partition of $\cZ = \cZ_1 \sqcup \cdots \sqcup \cZ_m$
\STATE \phantom{\bf Params:} $\triangleright$ Parameters $C$, $(C_\ell, \alpha_\ell)_{\ell \in [d]}$ with $\sum_{\ell=1}^d \alpha_\ell = 1$
\STATE \phantom{\bf Params:} $\triangleright$ Aggregatable Keys $K = ([d] \cup \{\bot\}) \times [m]$
\STATE {\bf Input:} Record $z \in \cZ$
\STATE {\bf Output:} Histogram contribution $w^z \in \Z_{\ge 0}^{K}$\\[-3mm]
\FOR{slice $j \in [m]$}
    \FOR{ $\ell \in [d]$}
    \STATE $w^z_{\ell,j} ~\gets~ \begin{cases}
                0 & \text{if } z \notin \cZ_j\\
                \CRR(q_\ell(z); C, C_\ell, \alpha_\ell) & \text{if } z \in \cZ_j
            \end{cases}$
    \ENDFOR
    \STATE $w^z_{\bot,j} ~\gets~ \begin{cases}
                0 & \text{if } z \notin \cZ_j\\
                \floor{\frac{\contribbudget}{C}} - \sum_{\ell=1}^d w^z_{\ell,j} & \text{if } z \in \cZ_j
            \end{cases}$
\ENDFOR
\RETURN $w^z$
\end{algorithmic}
\end{algorithm}

\begin{lemma}\label{lem:contribution-vector-norm}
For any $z$, the vector $w^z$ returned by \Cref{alg:contribution-vector-Linf} satisfies $w^z \ge \mathbf{0}$ and $\|w^z\|_1 = \floor{\contribbudget / C}$.
\end{lemma}
\begin{proof}
It is immediate to see that $\sum_{\ell,j} w^z_{\ell,j} + \sum_j w^z_{\bot,j} = \floor{\contribbudget / C}$. Let $j$ be such that $z \in \cZ_j$. To show that indeed all the values are non-negative, we observe that $w^z_{\ell,j'} = 0$ for all $j' \ne j$, and $0 \le \CRR(v; C, C_\ell, \alpha_\ell) \le \floor{\alpha_\ell \contribbudget / C}$ and hence, we have
\begin{align*}\textstyle
\sum_{\ell=1}^d w^z_{\ell,j}
~\le~ \sum_{\ell=1}^d \floor{\frac{\alpha_\ell \contribbudget}{C}}
~\le~ \floor{\frac{\contribbudget}{C} \sum_{\ell=1}^d \alpha_\ell}
~=~ \floor{\frac{\contribbudget}{C}}\,,
\end{align*}
and hence $w^z_{\bot,j} = \floor{\contribbudget/C} - \sum_\ell w^z_{\ell,j} \ge 0$.
\end{proof}

\noindent Thus, we have that for any impression $x$, at most the first $C$ conversions result in valid aggregatable reports.\pritish{Is there a need to elaborate more?}

To run through a concrete example, consider the \GiftShop dataset in \Cref{tab:dataset_example}, with queries $q_1(z)$ being the number of items purchased and $q_2(z)$ being the corresponding dollar value, and the slices corresponding to the ``Campaign''. In this case, the set of aggregatable keys $K$ is $\{1, 2, \bot\} \times \{\texttt{Thanksgiving}, \texttt{Christmas}\}$.
Suppose we use parameters $C = 2$, $C_1 = 2$, $C_2 = \$30$ and $\alpha_1 = \alpha_2 = 0.5$. Consider the record $z_1$ corresponding to Impression ID {\tt 123}. We have $v_1 = 3$ and $v_2 = \$21$, which get clipped as $v_1' = \clip_{C_1}(v_1) = 2$ and $v_2' = \clip_{C_2}(v_2) = \$21$. These get rescaled as $v_1'' = v_1' / C_1 = 1$ and $v_2'' = v_2' / C_2 = 0.7$. Note that $\alpha_1 \contribbudget / C = \alpha_2 \contribbudget / C = \contribbudget / 4 = 16384$. Thus, $z_1$ gets mapped to histogram contribution $w$ as
\begin{align*}
    w_{1, \texttt{Thanksgiving}} &~=~ \RR(16384 \cdot v_1'') 
    = \RR(16384) = 16384\\
    w_{2, \texttt{Thanksgiving}} &~=~ \RR(16384 \cdot v_2'') = \RR(11468.8) \stackrel{\mathrm{e.g.}}{=} 11469 \\
    w_{\bot, \texttt{Thanksgiving}} &~=~ 32768 - 16384 - 11469 = 4915\\
    w_{1, \texttt{Christmas}} &~=~ w_{2, \texttt{Christmas}} ~=~ w_{\bot, \texttt{Christmas}} ~=~ 0
\end{align*}
See \Cref{fig:hist_contrib} for an illustration of how this histogram contribution gets prepared at the browser/device.\pritish{I have cited \Cref{fig:hist_contrib} here. Should we move the figure also somewhere close by?}

It holds that $\|w^z\|_1 = 32768 = \contribbudget / 2$ for all $z$. Since records $z_1$, $z_2$, and $z_4$ correspond to the same impression $x$, the histogram contribution $w^{z_4}$ will end up being ignored in the aggregate, since $\|w^{z_1} + w^{z_2} + w^{z_4}\| > \contribbudget$. In other words, $\ocZ^x = (z_1, z_2)$.

\subsection{Reconstruction of Values}
Given the summary report $W$ (from \Cref{alg:summary-report}), we can post-process to obtain estimates of $V_D(q_\ell)$; see \Cref{alg:reconstruct}. We use $\ocZ^x_j$ to denote $\ocZ^x \cap \cZ_j$.

\begin{algorithm}[t]
\caption{\ReconstructValues $\cR$}
\label{alg:reconstruct}
\begin{algorithmic}
\STATE {\bf Params:} Queries $q_1, \ldots, q_d : \cZ \to \R_{\ge 0}$
\STATE \phantom{\bf Params:} Partition of $\cZ = \cZ_1 \sqcup \cdots \sqcup \cZ_m$.
\STATE {\bf Input:} Summary report $W \in \Z^{K}$ for $K = [m] \times ([d] \cup \{\bot\})$
\STATE {\bf Output:} $U \in \R^{(d+1) \times m}$ with $U_{k,j}$ corresponding to slice $j$, query $q_\ell$.
\FOR{slice $j \in [m]$}
    \FOR{ $\ell \in [d]$}
    \STATE $U_{\ell,j} ~\gets~ W_{\ell,j} \cdot C_\ell \Big / \floor{\frac{\alpha_\ell \contribbudget}{C}}$
    \ENDFOR
    \STATE $U_{0,j} ~\gets~ (W_{\bot,j} + \sum_{\ell \in [d]} W_{\ell,j}) \Big / \floor{\frac{\contribbudget}{C}}$
\ENDFOR
\RETURN $U$
\end{algorithmic}
\end{algorithm}

\begin{theorem}\label{thm:unbiased-estimate}
For $U$ returned by \Cref{alg:reconstruct}, it holds for all $\ell \in [d]$, that
\begin{align*}
    \Ex[U_{\ell, j}] &\textstyle~=~ \sum_{x \in X_j} \sum_{z \in \ocZ^x_j} \clip_{C_\ell}(q_\ell(z))\,,\\
    \text{and}\quad \Ex[U_{0, j}] &\textstyle~=~ \sum_{x \in X_j} |\ocZ^x_j|\,.
\end{align*}
Moreover, the variances are given as,
\begin{align*}
    \Var(U_{\ell, j}) &\textstyle~\le~ \prn{\frac{\sum_{x \in X_j} |\ocZ^x_j|}{4} + \Var(\DLap(\eps / \contribbudget))} \cdot C_\ell^2 \Big / \floor{\frac{\alpha_\ell \contribbudget}{C}}^2,\\
    \text{and}\quad \Var(U_{0, j}) &\textstyle~=~ \Var(\DLap(\eps / \contribbudget)) \cdot (d+1) \Big / \floor{\frac{\contribbudget}{C}}^2\,.
\end{align*}
\end{theorem}
\begin{proof}
It is easy to see that
\begin{align*}
    W_{\ell, j} &\textstyle~=~ \sum_{x \in X_j} \sum_{z \in \ocZ^x_j} \RR\prn{\floor{\frac{\alpha_\ell \contribbudget}{C}} \cdot \frac{\clip_{C_\ell}(q_\ell(z))}{C_\ell}} + \DLap\prn{\frac{\eps}{\contribbudget}}.
\end{align*}
Using the fact that $\Ex[\RR(\omega)] = \omega$ and $\Var(\RR(\omega)) \le \frac{1}{4}$, we get
\begin{align*}
    \textstyle\Ex[W_{\ell, j}] &\textstyle~=~ \sum_{x \in X_j} \sum_{z \in \ocZ^x_j} \floor{\frac{\alpha_\ell \contribbudget}{C}} \cdot \frac{\clip_{C_\ell}(q_\ell(z))}{C_\ell}, \\
    \textstyle\Var(W_{\ell, j}) &\textstyle~\le~ \sum_{x \in X_j} \frac{|\ocZ^x_j|}{4} + \Var\prn{\DLap\prn{\frac{\eps}{\contribbudget}}}.
\end{align*}
Similarly, we have that
\begin{align*}
    \textstyle\sum_{\ell=1}^d W_{\ell, j} + W_{\bot, j} &\textstyle~=~ \sum_{x \in X_j} \floor{\frac{\contribbudget}{C}} \cdot |\ocZ^x_j| + \sum_{\ell=0}^d \DLap\prn{\eps/\contribbudget},
\end{align*}
and hence
\begin{align*}
    \textstyle\Ex\sq{\sum_{\ell=1}^d W_{\ell, j} + W_{\bot, j}} &\textstyle~=~ \sum_{x \in X_j} \floor{\frac{\contribbudget}{C}} \cdot |\ocZ^x_j|, \\
    \textstyle\Var\prn{\sum_{\ell=1}^d W_{\ell, j} + W_{\bot, j}} &\textstyle~=~ (d+1) \cdot \Var(\DLap\prn{\eps/\contribbudget}).
\end{align*}
The proof is now complete by observing that $U_{\ell,j}$ and $U_{0,j}$ are simply scaled versions of $W_{\ell,j}$ and $W_{\bot,j} + \sum_{\ell=1}^d W_{\ell,j}$ respectively.
\end{proof}

\subsection{Optimization of Parameters}\label{subsec:optimization-params}

\begin{algorithm}[t]
\caption{\ParameterOptimization}
\label{alg:parameter-optimization}
\begin{algorithmic}
\STATE {\bf Params:} $\triangleright$ Queries $q_1, \ldots, q_d : \cZ \to \R_{\ge 0}$
\STATE \phantom{\bf Params:} $\triangleright$ Partition of $\cZ = \cZ_1 \sqcup \cdots \sqcup \cZ_m$.
\STATE {\bf Input:} Dataset $D \in \cZ^*$
\STATE {\bf Output:} Parameters $C$, $(C_\ell, \alpha_\ell)_{\ell \in [d]}$ with $\sum_{\ell=1}^d \alpha_\ell = 1$.
\STATE $R_{\mathrm{current}} \gets \infty$
\STATE $C_{\mathrm{\max}} \gets \max_{x\in X} |\cZ^x|$ \hfill \textcolor{black!60}{(max conversions for an impression)}
\FOR{$\widehat{C} = 1, 2, \ldots, C_{\mathrm{\max}}$}
    \STATE $(\widehat{C}_{\ell}, \widehat{\alpha}_\ell)_{\ell\in[d]} \gets \argmin_{(C_{\ell}, \alpha_\ell)_{\ell\in[d]}} R(\widehat{C}, (C_{\ell}, \alpha_\ell)_{\ell\in[d]})$
    \IF{$R(\widehat{C}, (\widehat{C}_{\ell}, \widehat{\alpha}_\ell)_{\ell\in[d]}) < R_{\mathrm{current}}$}
        \STATE $(C^*_{\ell}, \alpha^*_\ell)_{\ell\in[d]} \gets (\widehat{C}_{\ell}, \widehat{\alpha}_\ell)_{\ell\in[d]}$
        \STATE $C^* \gets C$
        \STATE $R_{\mathrm{current}} \gets R(\widehat{C}, (\widehat{C}_{\ell}, \widehat{\alpha}_\ell)_{\ell\in[d]})$
    \ENDIF
\ENDFOR
\RETURN $C^*, (C^*_{\ell}, \alpha^*_\ell)_{\ell\in[d]}$
\end{algorithmic}
\end{algorithm}

Having understood the variance in the estimates, we turn to the question of understanding the optimal choice of parameters $C$, $(C_\ell, \alpha_\ell)_{\ell \in [d]}$, with the goal of minimizing $\RMSRE_{\btau}(U, (q_\ell)_{\ell=0}^d; D)$. In this section, we optimize the choice of parameters, {\em using knowledge of the dataset $D$}. This is admittedly circular, as we are using an $\eps$-DP mechanism to learn information about a dataset {\em we do not know}. However, the eventual goal is that we will optimize the parameters using a historical dataset $D'$, which is ``similarly behaved'' to $D$ and as we show in \Cref{sec:generalization}, this is a reasonable choice as long as the two distributions are drawn from the same distribution.\pritish{Is last line okay?}

We stress that, since our optimization procedure is performed using the dataset $D'$ that is assumed to be public, this has no effect on the differential privacy guarantee of the sensitive dataset $D$.

We focus on minimizing $\RMSRE_{\btau}(U, (q_\ell)_{\ell=0}^d; D)^2$ as a function of $C, (C_{\ell}, \alpha_\ell)_{\ell\in[d]}$ given as
\begin{align*}
R(C, (C_{\ell}, \alpha_\ell)_{\ell\in[d]})^2 &\textstyle:= \frac{1}{(d+1)m} \sum\limits_{\ell=0}^d \sum\limits_{j \in [m]} \Ex \prn{\frac{U_{\ell,j} - V_D(q_\ell)_j}{\max\{\tau_{\ell}, V_D(q_\ell)_j\}}}^2.
\end{align*}
Denoting $\pi_{\ell,j} := 1 / \max\{\tau_{\ell}, V_D(q_{\ell})_j\}^2$, we can rewrite the objective as
\begin{align}
R(C, (C_{\ell}, \alpha_\ell)_{\ell\in[d]})^2 \textstyle:= \frac{1}{(d+1)m} \sum\limits_{\ell=0}^d \sum\limits_{j \in [m]} \pi_{\ell, j} \cdot \Ex (U_{\ell,j} - V_D(q_\ell)_j)^2.\label{eq:obj-simplified}
\end{align}
That is, $R$ is a linear combination of the following terms for $\ell \in \{0, \ldots, d\}$ and $j \in [m]$, for which we can use the bias-variance decomposition, namely
\begin{align*}
\Ex (U_{\ell, j} - V_D(q_\ell)_j)^2
&~=~ (V_D(q_\ell)_j - \Ex U_{\ell, j})^2 ~+~ \Var(U_{\ell, j}).
\end{align*}

\noindent Thus, we can choose the optimal parameters using a procedure as described in \Cref{alg:parameter-optimization}.
Namely, we enumerate over various values of $C$, and fixing $C = \hat{C}$, we optimize over the choice of $(C_\ell, \alpha_\ell)_{\ell \in [d]}$, and finally choose the value of $C$ and $(C_\ell, \alpha_\ell)_{\ell \in [d]}$ that minimizes $R(C, (C_\ell, \alpha_\ell)_{\ell \in [d]})$. The challenging step is the one computing $\argmin_{(C_{\ell}, \alpha_\ell)_{\ell\in[d]}} R(\widehat{C}, (C_{\ell}, \alpha_\ell)_{\ell\in[d]})$. In our experiments, we use the method \texttt{scipy.optimize.minimize}~\cite{scipyopt}, but in general any off-the-shelf optimizer could be used. In the worst-case, even if we do not minimize the objective exactly, it is still better than choosing the parameters in an ad hoc manner.

We show below that in fact $R(\widehat{C}, (C_{\ell}, \alpha_\ell)_{\ell\in[d]})$ is a non-convex objective in the parameters $(C_{\ell}, \alpha_\ell)_{\ell\in[d]}$, which can be hard to optimize in general. Nevertheless, we show that the objective is convex in $(C_\ell)_{\ell \in [d]}$ and $(\alpha_\ell)_{\ell \in [d]}$ separately. To recall,

\begin{definition}\label{def:convex}
A function $f : \R^t \to \R$ is {\em convex} if for all $x, y \in \R^t$ and $\lambda \in [0, 1]$, it holds that
$f(\lambda x + (1-\lambda) y) ~\le~ \lambda f(x) + (1-\lambda) f(y)\,.$
\end{definition}

\paragraph{Bias term.}
To simplify notation, let $\rem_{C_\ell}(v) := v - \clip_{C_\ell}(v) = \max\{0, v - C_\ell\}$. For $\ell \in [d]$, we have
\begin{align}
& V_D(q_\ell)_j - \Ex U_{\ell,j} \nonumber \\
&\textstyle~=~ \sum_{x \in X_j} \sum_{z \in \cZ^x_j} q_\ell(z) - \sum_{x \in X_j} \sum_{z \in \ocZ^x_j} \clip_{C_\ell}(q_\ell(z)) \nonumber \\
&\textstyle~=~ \sum_{x \in X_j} \sum_{z \in \cZ^x_j \smallsetminus \ocZ^x_j} q_\ell(z) + \sum_{x \in X_j} \sum_{z \in \ocZ^x_j} \rem_{C_\ell}(q_\ell(z))\nonumber\\
&\textstyle~=:~ B_{\ell,j}(C) + A_{\ell,j}(C_\ell, C)\label{eq:bias-simplified}
\end{align}

We observe that $\rem_{C_\ell}(q_\ell(z))$ is convex in $C_\ell$, and hence $A_{\ell,j}(C_\ell)$ is a convex function in $C_{\ell}$. Moreover, since $(V_D(q_\ell)_j - \Ex U_{\ell,j})$ is non-negative, we have the following, where we use the fact that the square of a non-negative convex function is convex.
\begin{observation}\label{obs:bias-convex}
$(V_D(q_\ell)_j - \Ex U_{\ell,j})^2$ is convex in $C_\ell$.
\end{observation}

\paragraph{Variance term.}
To simplify the optimization, we use the following relaxations in our calculations, that are obtained by
(i) approximating $\Var(\DLap(a)) \approx 2/a^2$ for $a \ll 1$, since we consider $\eps \ll \contribbudget$,
(ii) ignoring the variance due to randomized rounding,\footnote{The variance due to rounding in $W_{\ell,j}$ is at most $|X_j| \cdot C / 4$, which we view as much smaller than $\Var(\DLap(\eps/\contribbudget)) \approx (\contribbudget / \eps)^2$. This is reasonable because, e.g., if $\eps = 1$, then $(\contribbudget / \eps)^2 = 2^{32}$, which is typically order of magnitude larger than $|X_j| \cdot C / 4$ in practice.}\pritish{Please double check footnote!} and
(iii) approximating $\floor{\alpha_\ell \contribbudget / C}$ and $\floor{\contribbudget / C}$ as $\alpha_\ell \contribbudget / C$ and $\contribbudget / C$ respectively.
\begin{relaxation}\label{relax:estimate-variance}
We use the following approximations.
\[\textstyle
    \Var(U_{\ell, j}) ~\approx~ \frac{2 C^2 C_\ell^2}{\alpha_\ell^2 \eps^2} \ \text{ and } \ 
    \Var(U_{0, j}) ~\approx~ \frac{2(d+1)C^2}{\eps^2}\,.
\]
\end{relaxation}
\noindent Finally, we note that the function $a^2/b^2$ is non-convex in $(a, b)$. 

\begin{observation}\label{obs:var-convex}
Under \Cref{relax:estimate-variance}, for fixed $C \in \Z_{\ge 0}$ and $\ell \in [d]$, $\Var(U_{\ell, j})$ is
\begin{itemize}
\item convex in $C_{\ell}$ for a fixed $\alpha_\ell$.
\item convex in $\alpha_{\ell}$ for a fixed $C_\ell$.
\item non-convex in joint variables $(C_{\ell}, \alpha_{\ell})$.
\end{itemize}
\end{observation}

\paragraph{Putting it together.}
Thus, combining \Cref{obs:bias-convex} and \Cref{obs:var-convex}, we have the following.

\begin{theorem}\label{thm:convex}
For fixed $C \in \Z_{\ge 0}$, $\RMSRE_{\btau}(U, (q_\ell)_{\ell=0}^d; D)^2$ is
\begin{itemize}
    \item convex in $(C_{\ell})_{\ell \in [d]}$ for fixed $(\alpha_\ell)_{\ell \in [d]}$, and
    \item convex in $(\alpha_{\ell})_{\ell \in [d]}$ for fixed $(C_\ell)_{\ell \in [d]}$,
    \item non-convex in joint variables $(C_{\ell}, \alpha_{\ell})_{\ell \in [d]}$.\pritish{Actually, we need some more argument, as sum of a convex and non-convex function can be convex. Eg. $f = 2x^2$, $g = -x^2$; $f+g$ is convex.}
\end{itemize}
\end{theorem}

\subsection{\boldmath \texorpdfstring{$\ell_1$}{L1} version of \texorpdfstring{\HistogramContribution}{ContributionVector}}

Finally we present a variant of \HistogramContribution in \Cref{alg:contribution-vector-L1}, which uses $\ell_1$ clipping instead of $\ell_\infty$ clipping that is employed in \Cref{alg:contribution-vector-Linf}. The main intuition for this algorithm is that $\ell_1$ clipping results in less loss of signal, especially when the different query values are negatively correlated or only weakly correlated with each other. Below, we show that the histogram contributions generated this way respect the same $\ell_1$-norm constraint as \Cref{alg:contribution-vector-Linf}.\pritish{Is this location appropriate for this subsection?}

\begin{algorithm}[t]
\caption{\HistogramContribution ($\ell_1$ version)}
\label{alg:contribution-vector-L1}
\begin{algorithmic}
\STATE {\bf Params:} $\triangleright$ Queries $q_1, \ldots, q_d : \cZ \to \R_{\ge 0}$
\STATE \phantom{\bf Params:} $\triangleright$ Partition of $\cZ = \cZ_1 \sqcup \cdots \sqcup \cZ_m$.
\STATE \phantom{\bf Params:} $\triangleright$ Parameters $C$, $(C_\ell)_{\ell \in [d]}$.
\STATE {\bf Input:} Record $z \in \cZ$
\STATE {\bf Output:} Histogram contribution $w^z \in \Z_{\ge 0}^{K}$ for $K = ([d] \cup \{\bot\}) \times [m]$\\[-3mm]
\FOR{slice $j \in [m]$}
    \IF{$z \notin \cZ_j$}
        \STATE $(w^z_{1, j}, w^z_{2, j}, \ldots, w^z_{d, j}, w^z_{\bot, j}) \gets \mathbf{0}$
    \ELSE
        \STATE $\bv \gets \prn{\frac{q_1(z)}{C_1}, \ldots, \frac{q_d(z)}{C_d}} \in \R^d$
        \STATE $\bu \gets \frac{\bv}{\max\{1, \|\bv\|_1\}} \cdot \frac{\contribbudget}{C}$
        \STATE $(w^z_{1, j}, w^z_{2, j}, \ldots, w^z_{d, j}) \gets (\floor{u_1}, \ldots, \floor{u_d})$
        \STATE $w^z_{\bot, j} = \floor{\frac{\contribbudget}{C}} - \sum_{\ell=1}^d w^z_{\ell,j}$
    \ENDIF
\ENDFOR
\RETURN $w^z$
\end{algorithmic}
\end{algorithm}

\begin{lemma}\label{lem:contribution-vector-norm-L1}
For any $z$, the vector $w^z$ returned by \Cref{alg:contribution-vector-L1} satisfies $w^z \ge \mathbf{0}$ and $\|w^z\|_1 = \floor{\contribbudget / C}$.
\end{lemma}
\begin{proof}
It is immediate to see that $\sum_{\ell,j} w^z_{\ell,j} + \sum_j w^z_{\bot,j} = \floor{\contribbudget / C}$. Let $j$ be such that $z \in \cZ_j$. Clearly $w^z_{\ell,j'} = 0$ for all $j' \ne j$. We have
\begin{align*}\textstyle
\sum_{\ell=1}^d w^z_{\ell,j}
~\le~ \sum_{\ell=1}^d \floor{u_i}
~\le~ \floor{\sum_{\ell=1}^d u_i}
~=~ \floor{\frac{\contribbudget}{C}}\,,
\end{align*}
and hence $w^z_{\bot,j} = \floor{\contribbudget/C} - \sum_\ell w^z_{\ell,j} \ge 0$.
\end{proof}

While we do not analyze the error and generalization bounds for \Cref{alg:contribution-vector-L1},\pritish{As pointed by Avinash, $\alpha_\ell$'s are not being used.} we compare it in our experiments, using the same algorithm for reconstructing the estimates~(\Cref{alg:reconstruct}).\pritish{What $\alpha_\ell$'s does this algorithm use?}

\section{Experimental Evaluation} \label{sec:experiments}

\subsection{Setup}
We evaluate our algorithms on three real-world datasets and three synthetic datasets, which are described in more detail in the following sections. Each dataset is partitioned into a training set and a test set. 
For the real-world datasets, the partition is based on timestamps; for the synthetic data, separate training and test sets are generated independently from the data distribution. The training set is used to choose contribution budgets and clipping threshold parameters, and the error is evaluated on the test set. For the synthetic datasets, the training set is also used to choose a count limit $C$; for the real-world datasets only click-level or conversion-level data is available, so the count limit is set to $1$.

We compare our 
optimization-based algorithm to a simple \emph{baseline} approach. The baseline uses an equal contribution budget for each query, including a separate query for count, so that $\alpha_0=\cdots=\alpha_d = \frac{1}{d+1}$. The baseline uses a fixed quantile\pasin{What's the quantile?} of the training data to choose the clipping threshold $C_\ell$ for each query (as well as the count limit $C$, for the synthetic datasets). Note that to choose these thresholds, the baseline also requires access to training data.

For each dataset, we partition it into slices based on one or more impression features, and estimate multiple queries corresponding to each slice. For the real-world datasets we consider three queries for each slice, corresponding to the count and two additional conversion features depending on the dataset. For the synthetic datasets we consider two queries for each slice, corresponding to the count and a single conversion feature.

For the error metric $\RMSRE_\btau$, for each query $q_\ell$ we choose $\tau_\ell$ to be five times the median value of the query on the records of the training dataset. This ensures invariance of the error metric to rescaling the data, and allows us to combine the errors from features of different scales by taking $\btau = (\tau_0,\ldots,\tau_d)$.

\subsection{Real-World Datasets}

\paragraph{Criteo Sponsored Search Conversion Log (\CSSCL) Dataset~\citep{tallis2018reacting}}

This dataset consists of $15{,}995{,}634$\pritish{I think this number is conditioned on clicks. Should we mention that the dataset also has non-converting clicks, but that is not relevant for us?}
clicks obtained from a sample of $90$-day logs of live traffic from Criteo Predictive Search. Each point contains information on a user action (e.g., time of click on an ad) and a potential subsequent conversion (purchase of the corresponding product) within a $30$-day attribution window. We consider the following attributes: \texttt{partner\_id}, \texttt{product\_price} $\times$ \texttt{Sale}, and \texttt{SalesAmountInEuro}. 

\paragraph{Ad-tech Real Estate Dataset.}
This dataset consists of approximately $100{,}000$ real estate conversions from a 30-day period. 
We consider the following attributes: 
three known impression level features \texttt{F1}, \texttt{F2}, \texttt{F3}, and two unknown
conversion features
\texttt{Price} and
\texttt{Quantity}. 

\paragraph{Ad-tech Travel Dataset.}
This dataset consists of approximately $30{,}000$ travel conversions from a 30-day period. 
We consider the following attributes: 
three known impression level features \texttt{F1}, \texttt{F2}, \texttt{F3}, and two unknown
conversion features
\texttt{Price} and
\texttt{Quantity}.

\subsection{Synthetic Data}
The impact of various options to use the ARA can be evaluated by testing different configurations. However, such empirical evaluation would require access to a conversion dataset. Access to conversion datasets can be restricted and slow due to privacy concerns, or such data may not be available to practitioners. One way to address these difficulties is to use synthetic data that replicates the characteristics of real data that is bucketed by the summary reports in ARA.

In this context, we present a method for generating synthetic data through statistical modeling of actual conversion datasets. Initially, we performed an empirical analysis of these real conversion datasets to uncover relevant characteristics for ARA. Specifically, we examined the count and value distributions within these real conversion datasets. Subsequently, we designed a pipeline that employs the acquired distribution knowledge to create a realistic synthetic dataset, customizable by provided input parameters. In the following sections, we elaborate on the distributions as well as the process of generating data using this pipeline.

\subsubsection{Dataset Generation}\label{subsubsec:data-generation}

\newcommand{\campaignId}{\texttt{campaignId}}
\newcommand{\geography}{\texttt{geography}}
\newcommand{\productCategory}{\texttt{productCategory}}
\newcommand{\conversionType}{\texttt{conversionType}}

\begin{table}[h!]
\centering
\begin{tabular}{|c c c |} 
 \hline
 Name & Feature type & Side  \\ [0.5ex] 
 \hline
 \campaignId & Categorical(16) & \multirow{3}{*}{Impression}  \\ 
 \geography & Categorical(8) & \\
 \productCategory & Categorical(2) & \\
 \hline
 \conversionType & Categorical(5) & \multirow{2}{*}{Conversion}  \\
 \texttt{value} & $\in \R_{\ge 0}$ & \\
 \hline
\end{tabular}
\caption{Impression and conversion side features.}
\label{table:impression-dimensions}
\end{table}

Let us assume \GiftShop runs various Ad campaigns with the features shown in Table~\ref{table:impression-dimensions}. Records relevant to the ARA are outlined below:
\begin{enumerate}[leftmargin=*]
  \item \textbf{Impressions:}\label{step:one}
  For every display of an advertisement, an impression record is generated on the client side. For a specific key, e.g. `\campaignId=1 \& \geography=3 \& \productCategory=2', there could be a few or numerous impressions. Modeling the distribution of these impressions is the initial aspect to address.
  \item \textbf{Conversions:} \label{step:two} An impression might lead to zero, one, or multiple conversion events. These conversion events are defined within the ad-tech context and encompass various activities, such as \textit{click},  \textit{add-to-cart},  \textit{purchase},  \textit{spend-30-seconds},  and \textit{achieved-level-2} to provide a few examples. So the next aspect to model is the count of conversions per impression, as well as the conversion features associated to it, such as \conversionType.
  \item \textbf{Value Contributions:} \label{step:four} Not every conversion yields the same return for advertisers. For instance, a purchase of \$25 might be more desirable than one of \$5. Beyond simply considering the number of conversions, it is crucial to take into account the value that these conversions generate. This leads us to the distribution of conversion values, which captures this aspect.
\end{enumerate}

\begin{figure}
\centering
\includegraphics[width=\columnwidth]{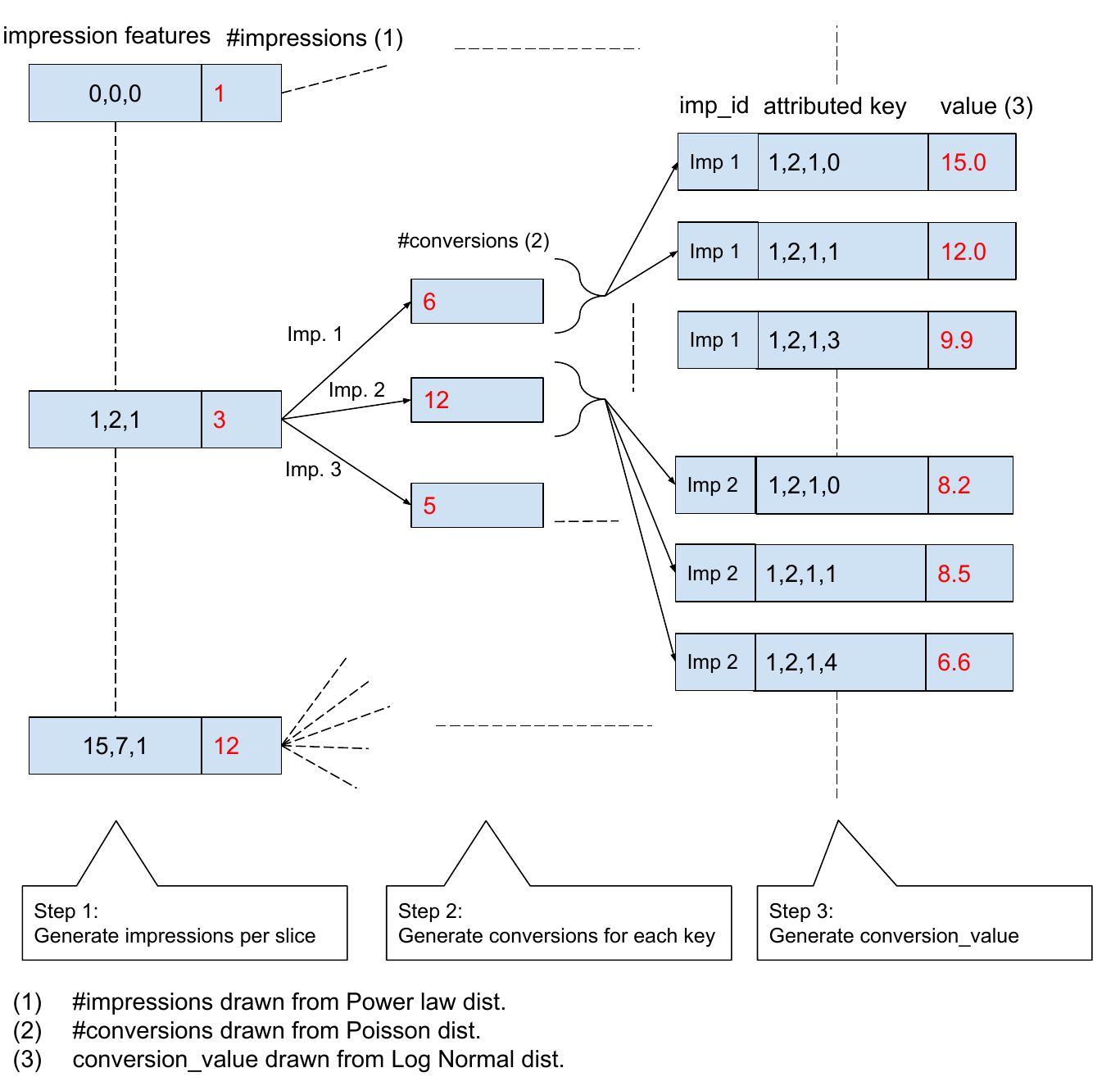}
\caption{Overall dataset generation steps with features in \Cref{table:impression-dimensions} used for illustration.}
\label{fig:pipeline}
\end{figure}

We propose a pipeline that generates both counts and values as shown in Figure~\ref{fig:pipeline}. Here, the data is not aggregated so that event level processing such as count bounding and contribution budgeted could be performed. First, we define the distributions that are used in the pipeline.
\begin{definition}[Power Law Distributions] \label{def:powerlaw}
The \emph{Power Law distribution} with parameter $b > 0$ is the distribution supported on positive integers, whose probability mass function at integer $k$ is
$$\Pr[X = k] ~=~ 
\begin{cases}
  \frac{k^{-b}}{\sum_{k=k_{\min}}^{k_{\max}} k^{-b}} & \text{} k_{\min} \le k \le k_{\max} \\
  0 & \text{otherwise,} 
\end{cases}
$$
where $b$ is shape parameter, $k_{\min}>0$ is lower bound and $k_{\max}$ is upper bound.
\end{definition}

A well-behaved distribution typically exhibits parameters of $1<b<3$, $k_{\min}=1$ and $k_{\max}=\infty$. However, when working with real datasets, it is common to observe power-law behavior within a specific range and arbitrary $b$ parameter.



\begin{definition}[Poisson Distribution]\label{def:poisson}
The \emph{Poisson distribution} with parameter $\lambda>0$ is a discrete probability distribution whose probability mass function at integer $k$ is
    \[P(X = k) =  \frac{\lambda^k e^{-\lambda}}{k!},\]
where the parameter $\lambda$ is the average rate of events.
\end{definition}


\begin{definition}[Log-Normal Distribution]\label{def:lognormal}
The \emph{Log-Normal distribution} with parameters  $\mu$ and $\sigma$ is a continuous probability distribution whose probability density function is
$$f(x) = \frac{1}{\sigma x \sqrt{2 \pi}} \exp \left( -\frac{(\log x - \mu)^2}{2 \sigma^2} \right).$$
\end{definition}

\begin{figure*}[tb]
\centering
\includegraphics[width=4in]{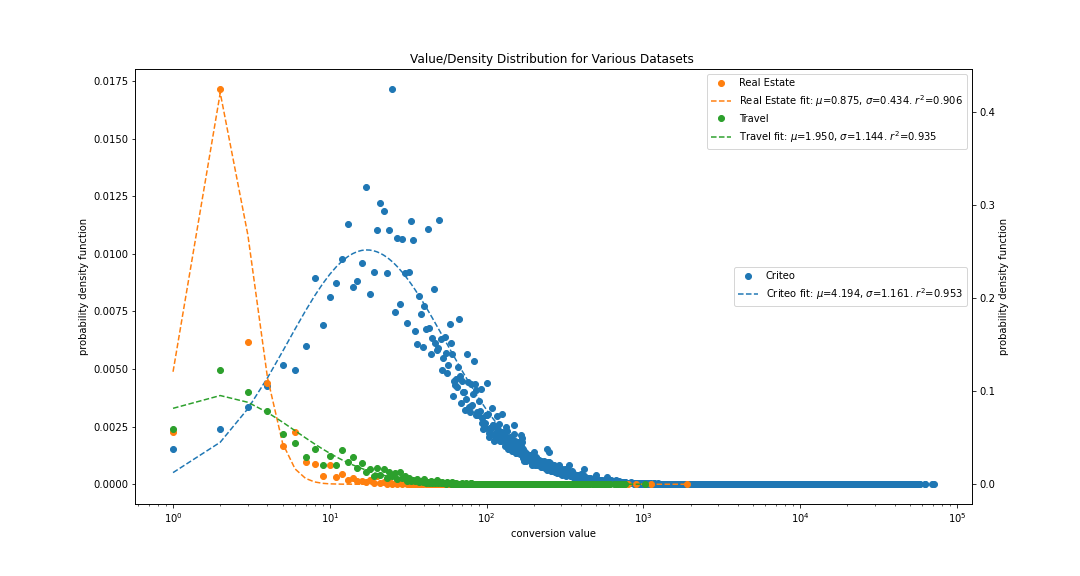}
\caption{Conversion value fits on the three real ads datasets.}
\label{fig:value-fit}
\end{figure*}

Having defined the relevant distributions, we now describe the data generation in more detail. For convenience of the theoretical analysis in the next section, we will describe the data generation for a general choice of distributions:
\begin{itemize}[leftmargin=*]
\item $\cDc$: the distribution of number of impressions per slice. In our experiments, this is set to the power-law distribution (with pre-specified parameter $b$). 
\item $\cDcc$: the distribution of the number of conversions per impression. In our experiments, this is set to the Poisson distribution (with pre-specified parameter $\lambda$). The conversions are subsequently divided uniformly among the different values of attributed keys.
\item $\cDv$: the distribution of conversion values. In our experiments, this is set to the Log-normal distribution (with pre-specified parameter $\mu, \sigma$). Figure~\ref{fig:value-fit} displays conversion values extracted from three datasets alongside the corresponding Log-Normal distribution fit.
\end{itemize}


\noindent Our data generation pipeline works in the following stages:
\begin{enumerate}
    \item [Step 1] For each combination of impression features, sample \#impressions is independently sampled from the distribution $\cDc$\footnote{Sampling from discrete power-law distributions with arbitrary parameters $b$ is not a straightforward process. To address this challenge, we adopted the approximation method outlined in Appendix D of the work by Clauset et al. in~\cite{clauset2009power}}.
    There will be \[T=\prod_{\cX_i \in \cX} |\cX_i|\] attributed slices, where $\cX$ represents the set of dimensions within the impression side. To illustrate, in the sample case shown in \Cref{table:impression-dimensions}, there will be $T = 16 \times 8 \times 2$ slices.
    \item [Step 2] For each impression, independently sample \#conversion from the distribution $\cDcc$, and distribute each one uniformly at random between the various conversion features. In the case of \Cref{table:impression-dimensions}, there are $5$ values of \conversionType.
    \item [Step 3] For each conversion, independently sample the conversion value from the distribution $\cDv$.
\end{enumerate}

\begin{table}[h!]
\centering
\begin{tabular}{|c|c|c|cc|} 
 \hline
 \multirow{2}{*}{Name} & Step 1 & Step 2 & \multicolumn{2}{c|}{Step 3} \\ [0.5ex] 
 \cline{2-5}
 & $b$ & $\lambda$ & $\mu$ & $\sigma$ \\
 \hline
 {\tt synth-criteo} & $2.88$ & $10$ & $4.19$ & $1.16$  \\ 
 {\tt synth-real-estate} & $0.06$ & $10$ & $0.87$ & $0.43$  \\ 
 {\tt synth-travel} & $1.14$ & $10$ & $1.95$ & $1.14$  \\ 
 \hline
\end{tabular}
\caption{Synthetic datasets utilized in evaluations with parameters that mimic the corresponding real datasets.}

\label{table:synthetic-datasets}
\end{table}

Table~\ref{table:synthetic-datasets} presents three synthetic datasets that were employed in evaluations with parameters. It is possible to generate numerous datasets with specific parameters that closely mimicking the characteristics of a target dataset. This could be particularly useful for emulating privacy-restricted proprietary ad datasets.

\def\figwidth{0.31}
\def\figspacing{1mm}
\begin{figure*}[tb]
  \centering
  \subfigure[Criteo sponsored search conversion log]{
    \includegraphics[width=\figwidth\linewidth]{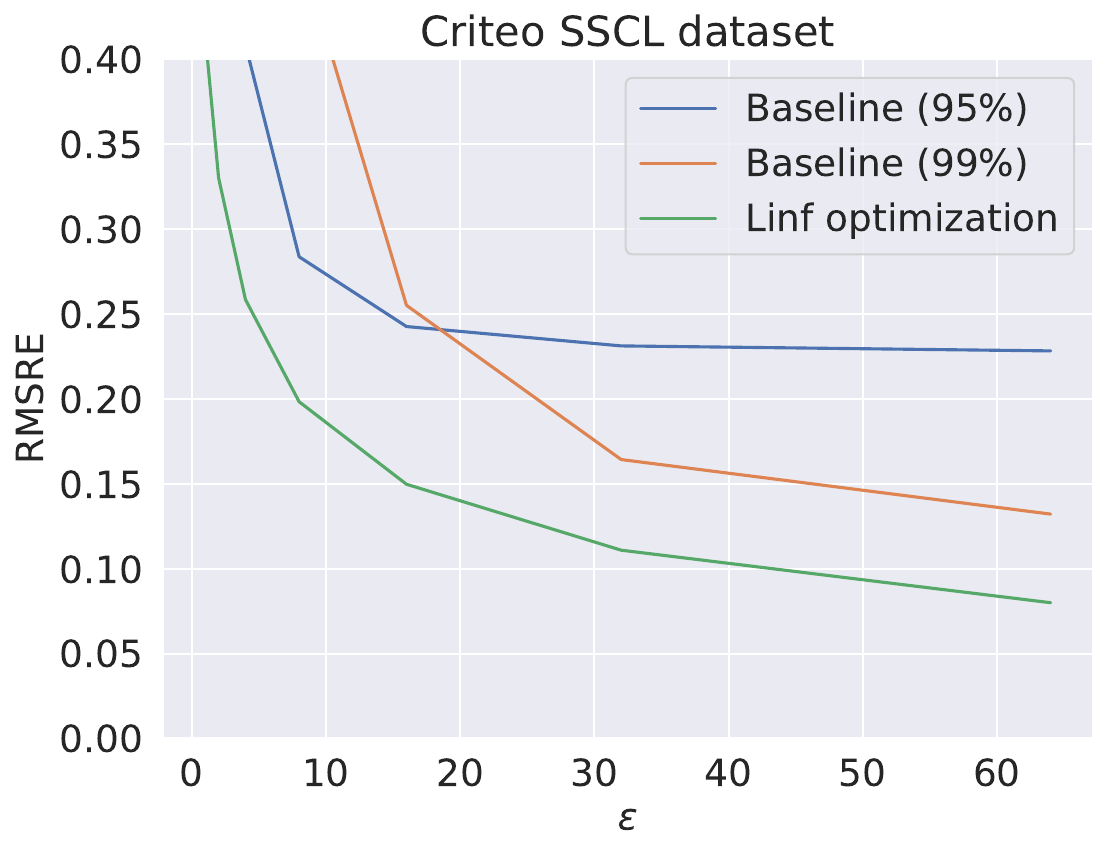}
    \label{fig:expt_criteo}
  }
  \hspace{\figspacing}
  \subfigure[Ad-tech real estate]{
    \includegraphics[width=\figwidth\linewidth]{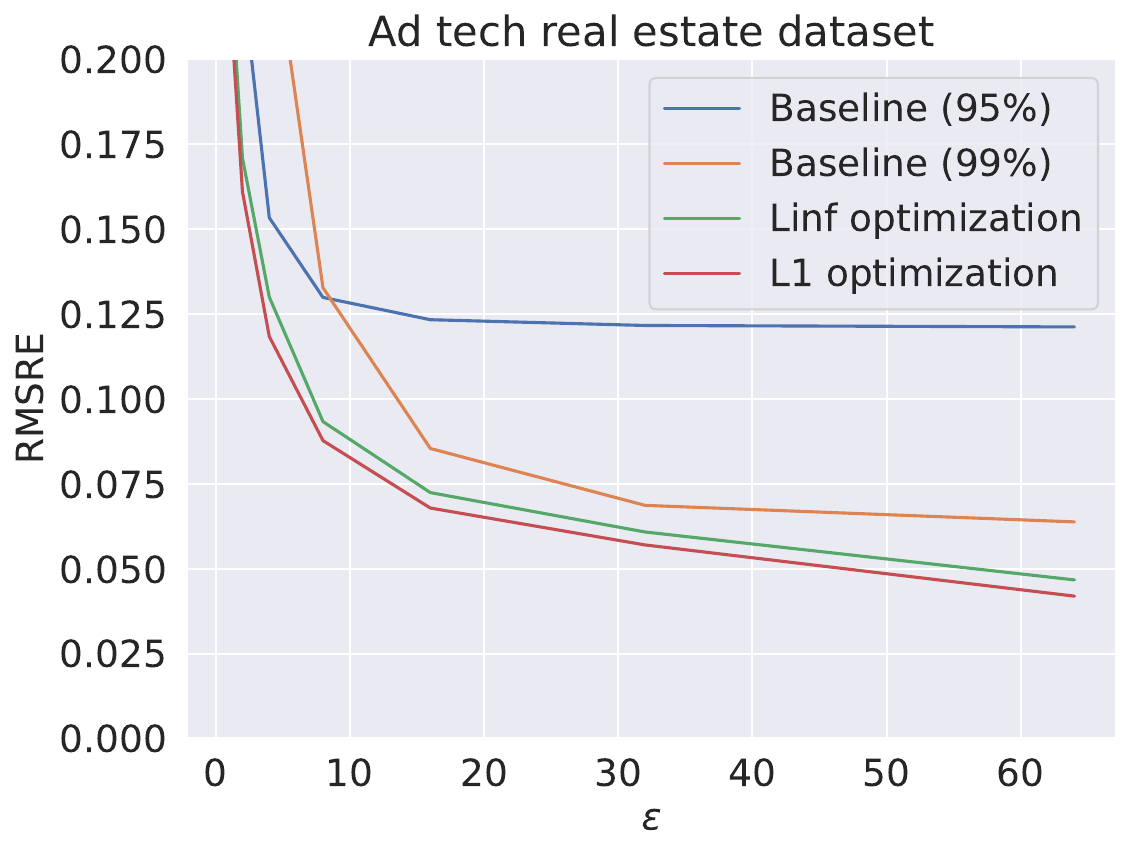}
    \label{fig:expt_re}
  }
  \hspace{\figspacing}
  \subfigure[Ad-tech travel]{
    \includegraphics[width=\figwidth\linewidth]{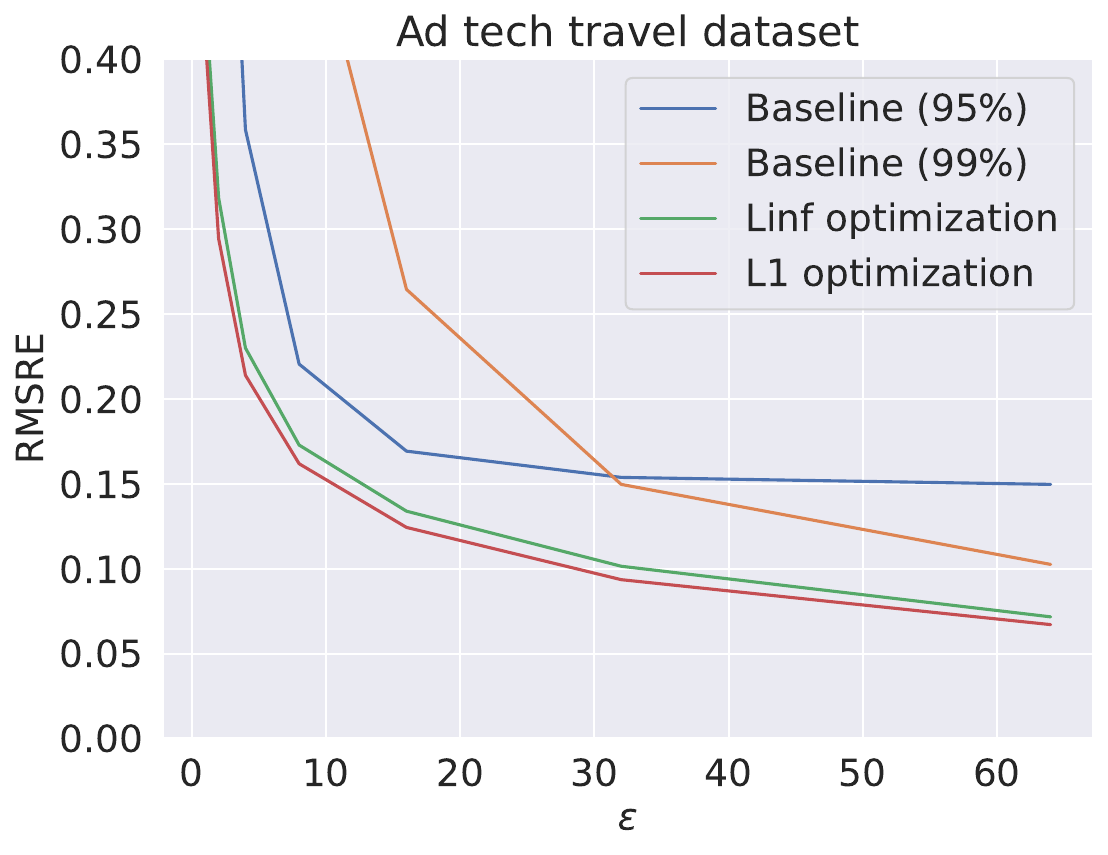}
    \label{fig:expt_travel}
  }

  \subfigure[synth-criteo]{
    \includegraphics[width=\figwidth\linewidth]{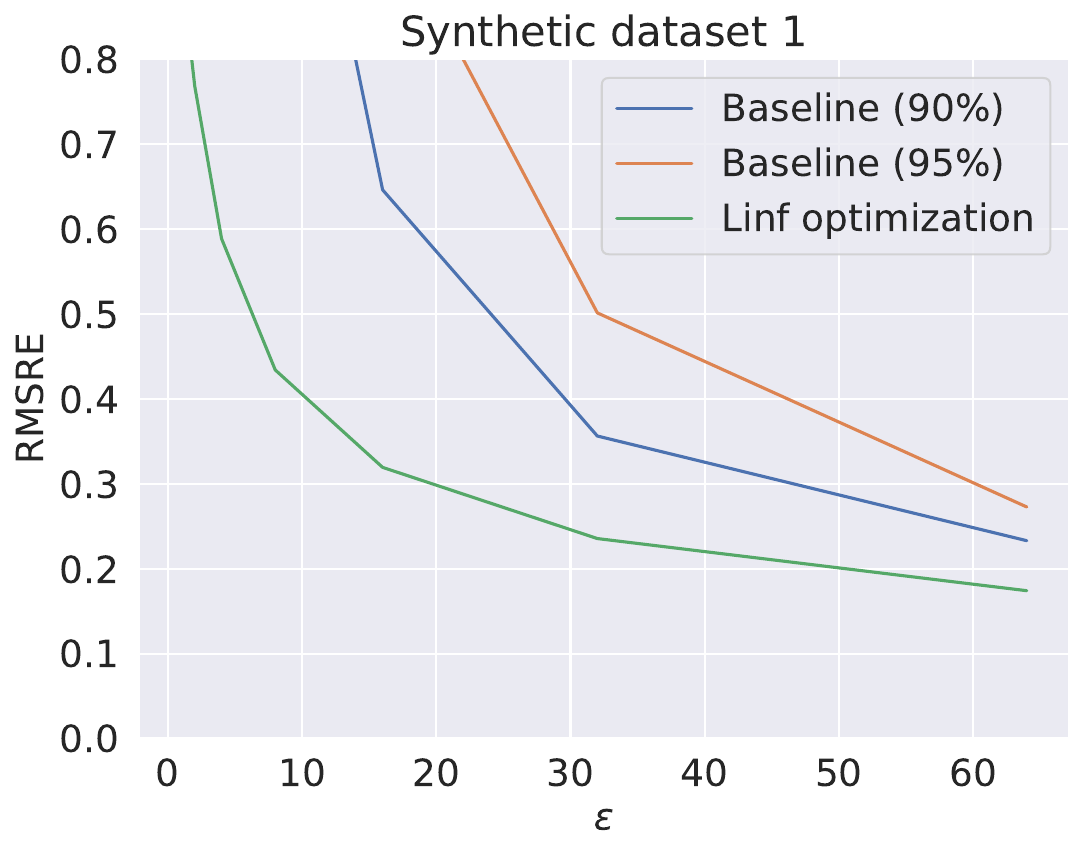}
    \label{fig:expt_synth1}
  }
  \hspace{\figspacing}
  \subfigure[synth-real-estate]{
    \includegraphics[width=\figwidth\linewidth]{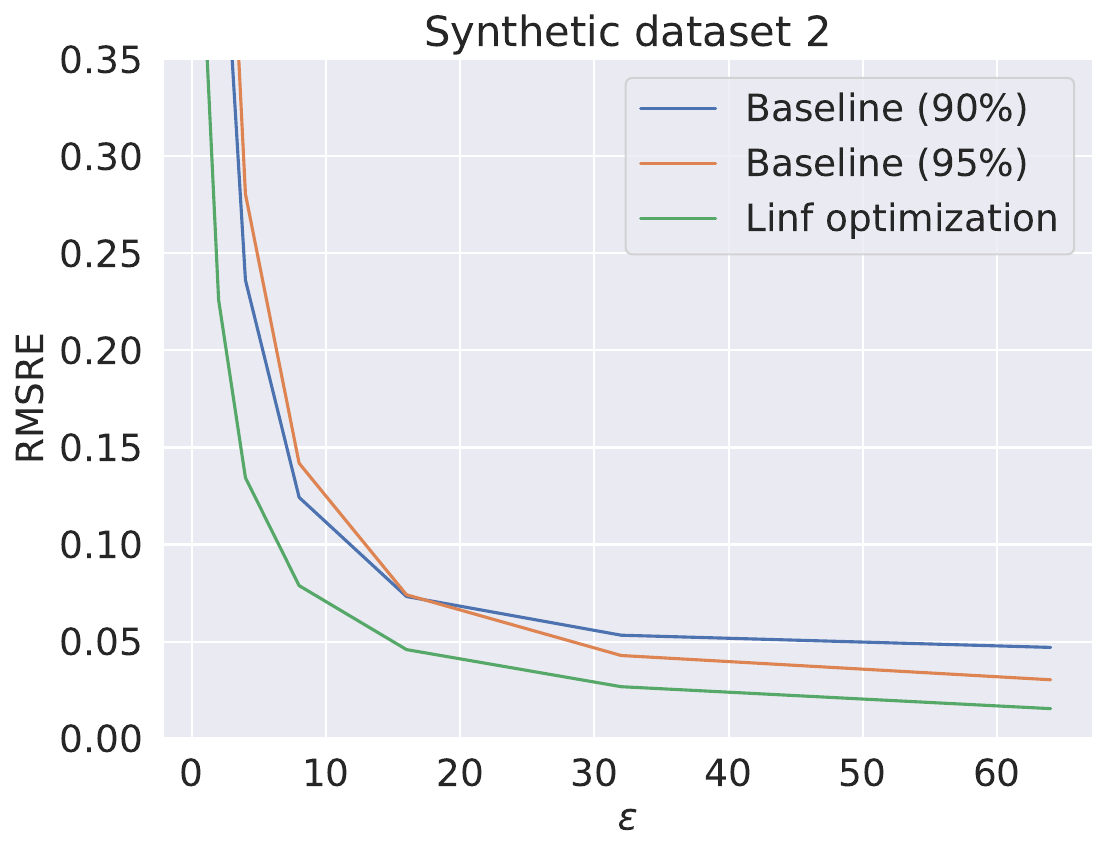}
    \label{fig:expt_synth2}
  }
  \hspace{\figspacing}
  \subfigure[synth-travel]{
    \includegraphics[width=\figwidth\linewidth]{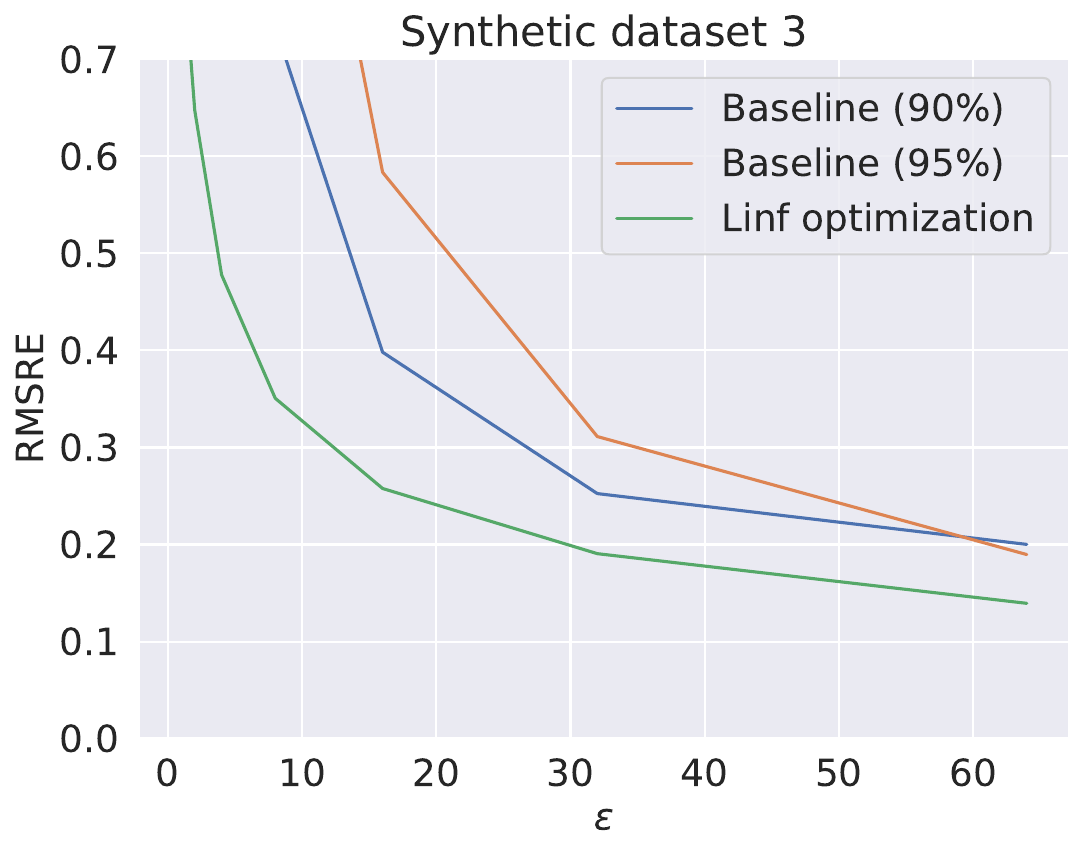}
    \label{fig:expt_synth3}
  }

  \caption{\boldmath $\RMSRE_\tau$ for privacy budgets $\{1,2,4,8,16,32,64\}$ for our algorithms and baselines on three real-world and three synthetic datasets.}
  \label{fig:experiments}
\end{figure*}

\subsection{Results}

We see in Figure~\ref{fig:experiments} that the estimates produced by our algorithms have substantially lower error than the baselines, on both the real-world and synthetic datasets. Moreover, the excess error incurred by each baseline depends on the data and overall privacy budget. In contrast, our optimization-based approach is able to adapt to the privacy budget and data.

For the real-world real estate and travel datasets, we additionally compare our $\ell_1$ optimization-based algorithm and show that it provides additional accuracy improvements. The $\ell_1$ approach provides only a marginal improvement on the criteo dataset, likely due to high correlation between the queried features, and so we do not include it in the plot. For the synthetic datasets the $\ell_1$ optimization is equivalent to $\ell_\infty$ optimization, since there is only a single non-count query on these datasets.  


\section{Generalization Bounds}\label{sec:generalization}

Since we optimize the parameters on the historical dataset, it is important to ensure that we are not overfitting to this training dataset in such a way that it performs badly on the actual (i.e.,  test) dataset. To support our empirical findings, in this section, we formally prove---in a simplified setting---a generalization bound showing that the expected RMSRE on the actual dataset is close to optimal even with this procedure.

We work in the data generation model as in the previous section. 
For the purpose of theoretical analysis,  we consider a simplified setting where: (i) there is only one conversion per impression (i.e., $\cDcc$ is the point-mass distribution that is always equal to one) and (ii) that there is only a single query (i.e., $d = 1$). 

Recall the notations from \Cref{sec:capping}.
Due to (i), we always set the per-impression count capping to $C = 1$; this also gives $\cZ^x = \ocZ^x$ for all impression $x$. For convenience, we also define the following notations: 
\begin{align*}
\bias_{C_1}(D_j) &:= \sum_{x \in X_j} \sum_{z \in \cZ^x} \rem_{C_1}(q_1(z)_j), \\
\numabove_{C_1}(D_j) &:= \sum_{x \in X_j} \sum_{z \in \cZ^x} \ind[q_1(z)_j > C_1], \\
\pi(D_j) &:= \max\{\tau_1, V_D(q_1)_j\}, \\
R_{D_j}(C_1) &:= \frac{1}{\pi(D_j)^2} \prn{\bias_{C_1}(D_j)^2 + \frac{2C_1^2}{\eps^2}},
\end{align*}
where recall that $X_j$ is the set of all $x$ such that $(x,y) \in D_j$ for some $y$, and $R$ is similar to \Cref{eq:obj-simplified}, but here, we only have one argument, namely $C_1$. $\bias_{C_1}(D_j)$ is the {\bf bi}as incurred in the estimate due to clipping; this is similar to the term in \Cref{eq:bias-simplified}, and $\numabove_{C_1}(D_j)$ counts the {\bf n}umber of conversion values that were {\bf c}lipped by the threshold at $C_1$.
As stated earlier, we assume that the number of impressions in the $j$th slice is generated by $\cDc$ (with one conversion per impression) and the value of each conversion is generated independently by $\cDv$. We denote this entire compound distribution by $\cDcom$.

Finally, we let
\begin{align*}
\tR_{\cDcom}(C_1) := \E_{D \sim \cDcom}\left[R_D(C_1)\right],
\end{align*}
denote the expected loss $R_D(C_1)$ where $D$ is drawn from the distribution $\cDcom$.

In this simplified setting, the optimization objective reduces to just minimizing
\begin{align*}
R_{D_1, \dots, D_m}(C_1)
&:= \frac{1}{m}\prn{R_{D_1}(C_1) + \cdots + R_{D_m}(C_1)} \\
&= \frac{1}{m} \sum_{j \in [m]} \frac{1}{\pi(D_j)^2} \prn{\bias_{C_1}(D_j)^2 + \frac{2C_1^2}{\eps^2}}.
\end{align*}
Differentiating this (w.r.t. $C_1$), we get
\begin{align*}
&\frac{\partial}{\partial C_1} R_{D_1, \dots, D_m}(C_1)\\
&= \frac{1}{m} \sum_{j \in [m]} \frac{1}{\pi(D_j)^2} \prn{\frac{4C_1}{\eps^2} - 2\bias_{C_1}(D_j)\cdot \numabove_{C_1}(D_j)}.
\end{align*}
In other words, the optimum contribution bounding threshold $C^*_1 = C^*_1(D_1, \dots, D_j)$ is such that
\begin{align*}
\sum_{j \in [m]} \frac{1}{\pi(D_j)^2} \prn{\frac{4C^*_1}{\eps^2} - 2\bias_{C^*_1}(D_j) \cdot \numabove_{C_1}(D_j)} = 0.
\end{align*}

Let $\mu_n(\cD)$ denote the $n$th moment of the distribution $\cD$ over $\R$, namely, $\mu_n(\cD) = \Ex_{x \sim \cD} |x|^n$. We can get the following generalization bound. Note that the LHS is the expected error of the fresh (independent) slice if we optimize based on the $D_1, \dots, D_m$ (i.e., historical data) drawn from the same distribution $\cDcom$, while the RHS is the expected error with respect to the optimal threshold for the distribution.

\begin{theorem}
For any distributions $\cDc, \cDv$ such that the moments $\mu_4(\cDc)$ and $\mu_2(\cDv)$ are finite. For any $\zeta, \theta > 0$, there exists $m \in \N$ such that, with probability $1 - \zeta$ over $D_1, \dots, D_m \sim \cDcom$, and using $C_1^* = C_1^*(D_1, \ldots, D_m)$, we have
\begin{align*}
\tR_{\cDcom}(C^*_1) ~\leq~ \min_{\tC_1 \geq 0} \tR_{\cDcom}(\tC_1) + \theta.
\end{align*}
\end{theorem}

\begin{proof}
Let $\tC^*_1 = \argmin_{\tC_1 \geq 0} \tR_{\cDcom}(\tC_1)$ denote the optimal clipping threshold of the distribution. Since $\tR_{\cDcom}(\tC_1)$ is a continuous function, there exists $\lambda > 0$ such that
\begin{align} \label{eq:neighborhood-loss}
\tR_{\cDcom}(C_1) - \tR_{\cDcom}(\tC^*_1) \leq \theta. & &\forall C_1 \in [\tC^*_1 - \lambda, \tC^*_1 + \lambda].
\end{align}
Furthermore, let $\nu := \E[\frac{1}{\pi(X)^2}] > 0$. 

Recall that
\begin{align*}
\frac{\partial}{\partial C_1} \tR_{X}(\tC^*_1) = \frac{1}{\pi(X)^2} \prn{\frac{4\tC^*_1}{\eps^2} - 2\bias_{\tC^*_1}(X) \cdot \numabove_{\tC^*_1}(X)}.
\end{align*}
Note that $\E_{X \sim \cDcom}\sq{\frac{\partial}{\partial C_1} \tR_X(\tC^*_1)} = 0$ (due to $\tC^*_1$ being the minimizer).
Furthermore, we have
\begin{align*}
&\E_{X \sim \cDcom}\sq{\prn{\frac{\partial}{\partial C_1} \tR_{X}(\tC^*_1)}^2}\\
&\leq O\prn{\frac{1}{\pi^2} \prn{\frac{(\tC^*_1)^2}{\eps^4} + \bias_{\tC^*_1}(X)^2 \cdot \numabove_{\tC^*_1}(X)^2}}  \\
&\leq O\prn{\frac{1}{\pi^2} \prn{\frac{(\tC^*_1)^2}{\eps^4} + \bias_{0}(X)^2 \cdot \numabove_{0}(X)^2}}  \\
&= O\prn{\frac{1}{\pi^2} \prn{\frac{(\tC^*_1)^2}{\eps^4} + \mu_2(\cDv)\mu_4(\cDc)}},
\end{align*}
which is finite under the assumption in the theorem statement.
Thus, $\frac{\partial}{\partial C_1}\tR_{X}(\tC^*_1)$ when $X \sim \cDcom$, $\tR'_{X}(\tC^*_1)$ has a finite variance. Similarly, the term $\frac{1}{\pi(X)^2}$ has a finite variance, simply because its maximum value is at most $1/\tau_1^2$. Thus, for any $\zeta, \theta > 0$, there exists $m_0$ such that for any $m \geq m_0$, with probability $1 - \theta$ over $D_1, \dots, D_m \sim \cDcom$, both of the following hold:
\begin{align*}
\pabs{\frac{\partial}{\partial C_1} R_{D_1,\dots,D_m(C_1)}} = \pabs{\frac{1}{m} \sum_{j \in [m]} \frac{\partial}{\partial C_1} R_{D_j}(C^*_1)} \leq 4\lambda\nu / \eps^2,
\end{align*}
and
\begin{align*}
\frac{1}{m} \sum_{j \in [m]} \frac{1}{\pi(D_j)^2} \geq \nu/2.
\end{align*}
Now, notice that the objective $R_{D_1, \dots, D_m}(C_1)$ is $\prn{\frac{4}{m \eps^2} \sum_{j \in [m]} \frac{1}{\pi(D_j)^2}}$-strongly convex.\pritish{Define strong-convexity.} As a result, when the above two inequalities hold we have
\begin{align*}
|\tC^*_1 - C^*_1| \leq \frac{\pabs{\frac{\partial}{\partial C_1} R_{D_1,\dots,D_m}(C_1)}}{\prn{\frac{4}{m \eps^2} \sum_{j \in [m]} \frac{1}{\pi(D_j)^2}}} \leq \lambda.
\end{align*}
From \eqref{eq:neighborhood-loss}, this implies $\tR_{\cDcom}(C^*_1) - \tR_{\cDcom}(C^*_1)$.
\end{proof}

We remark that, due to the use of continuity argument of $\tR_{\cDcom}$ (at $\tC^*_1$), we do not achieve any explicit bound in the rate of convergence. It remains an interesting question to extend this argument to get a specific rate. Similarly, it remains interesting to incorporate the privacy budgets (i.e., $\alpha_\ell$'s) in the presence of multiple queries to the bounds as well.

\section{Related Work}\label{sec:related_work}

The work closest to ours is that  on optimizing hierarchical queries when using the Attribution Reporting API \cite{dawson2023optimizing}.\pritish{Not sure if we should elaborate more on the definition of the hierarchical setting.} In our terminology, this corresponds to aggregating with respect to multiple partitions of $\cZ$ that are refinements of each other; our work is complementary in that we focused on a single partition. In addition, our setting involves aggregating different conversion values for each slice.  Thus, while~\cite{dawson2023optimizing} optimized for contribution budget allocation across different slices, our work optimizes the contribution budget allocation across the different queries for each slice.  Furthermore,~\cite{dawson2023optimizing} also involved post-processing the estimates that ensured consistency of estimates and reduced the overall noise; this was done by generalizing the methods in~\cite{hay2009boosting,cormode2012differentially}. Such post-processing is not relevant in our context as we do not have any consistency constraints that are satisfied by the noiseless data.  Hence, it is possible to combine the techniques in our work with the techniques in \cite{dawson2023optimizing} to consider a setting where we have hierarchical queries with multiple conversion values to aggregate.

Private aggregation by contribution bounding and adding noise is a common technique in DP. It was shown in \cite{amin2019bounding} that in order to minimize the $\ell_1$-error, the optimal threshold is to set the contribution bound to be the $(1 - 1/\eps n)$th percentile of the data.\pritish{What happens when $\eps > 1$?} On the other hand, in order to minimize the $\ell_2^2$-error, it was shown in \cite{kamath2023biasvarianceprivacy} that bounding the range and adding Laplace noise achieves the smallest error, thereby establishing a bias-variance-privacy trilemma; this is precisely what we get in our approach as well, where we clip the value range and add (discrete) Laplace noise, by optimizing the clip threshold using historical data.\pritish{Anything more relevant to say here?}

\section{Conclusion and Future Directions}\label{sec:conc_future_directions}

In this work, we studied the optimization of summary reports in the ARA, which is currently deployed on hundreds of millions of Chrome browsers. To the best of our knowledge, there has been no prior work formulating the contribution budgeting optimization problem for ARA. We hope that our rigorous formulation will equip researchers with the right abstraction of the problem as well as the API to develop DP  algorithms for ad conversion measurement with better privacy-utility trade-offs.

Our recipe, which leverages past data that is noiseless and that has not been bounded, in order to bound the contributions in future data when querying it with DP, is quite general and applicable to settings (beyond advertising) where a system is queried continuously over time, and a DP constraint is being continuously enforced. We note that one approach based on this work is to learn the parameter(s) of the synthetic data distribution using past data, and then sample repeatedly from this distribution to construct a synthetic dataset that can be used for privacy budgeting for queries on future data. 


Another very interesting direction for future work is to develop algorithms that do not rely on non-contribution bounded noiseless data for optimizing the contribution bounding parameters used for querying future data. While such non-contribution bounded noiseless data might still be available for long-running campaigns, new campaigns launched well after the deprecation of third-party cookies would benefit from methods for continuously updating the contribution bounding parameters based solely on the outputs of privacy-preserving APIs.

In addition to summary reports, ARA offers \emph{event-level reports} \cite{event-api-android} which are also subject to (a different type of) contribution bounding and noising; our method does not take these reports into account when setting the contribution bounds for summary reports. It would be interesting to explore whether event-level reports can be leveraged to optimize the summary reports in ARA.

As described in \Cref{sec:ara_summary_reports_constraints}, summary reports in ARA are currently restricted by on-client attribution and by the \emph{separate} computation of the contributions of different attributed conversions. It would be interesting to determine the utility improvement that could be achieved if the contributions of different attributed conversions can be computed \emph{jointly}, e.g., if attribution were to be done off-client either in a trusted execution environment or via a secure multi-party computation protocol, or alternatively if the contributions of an attributed conversion can simply take into account the contributions of previously attributed conversions on the same client. 


\bibliographystyle{acm}
\bibliography{refs}

\newcommand{\ARE}{\mathsf{ARE}}
\newcommand{\AME}{\mathsf{AME}}
\newcommand{\APME}{\mathsf{APME}}
\newcommand{\EARE}{\mathsf{EARE}}
\newcommand{\REARE}{\mathsf{REARE}}
\newcommand{\NSR}{\mathsf{NSR}}
\newcommand{\EAREO}{\mathsf{EAREO}}

\begin{table*}[ht]
\renewcommand{\arraystretch}{1.7}
\centering
\begin{tabular}{lcll}
\toprule
Short Name & Error Metric (slice $j$) & Sample Parameters & Interpretation \\
\midrule 
$\ARE_\alpha$ & $\Pr\sq{\frac{|U_{j}-V(q)_{j}|}{V(q)_{j}}>\alpha}$ & $\alpha \in \{0.1, 0.2\}$ & Probability of seeing large relative error. \\
$\AME_\tau$ & $\Pr\sq{|U_{j}-V(q)_{j}|> \tau}$ & $\tau \in \{1,5\}$ & Probability of seeing large magnitude errors. \\
$\APME_{\alpha, \tau}$ & $\Pr\sq{{\frac{|U_{j}-V(q)_{j}|}{V(q)_{j}}>\alpha} \cap {|U_{j}-V(q)_{j}|>\tau}}$ & $\alpha\in\{0.2\}, \tau\in \{1, 5\}$ & Probability of seeing large magnitude and relative errors. \\
$\EARE$ & $\Ex\sq{\frac{|U_{j}-V(q)_{j}|}{V(q)_{j}}}$ &   & Expected absolute relative error (to true value).\\
$\RMSE$ & $\sqrt{\Ex \sq{\prn{U_{j}-V(q)_{j}}^2}}$ &  & Root mean squared error. \\
$\RMSRE$ & $\sqrt{\Ex \sq{\prn{\frac{U_{j}-V(q)_{j}}{V(q)_{j}}}^2}}$ &  & Root mean squared relative error. \\
$\EARE_\tau$ & $\Ex\sq{\frac{|U_{j}-V(q)_{j}| }{ \max(\tau, V(q)_{j})}}$  & $\tau\in\{3, 5,10\}$ & Mean absolute relative error at threshold $\tau$. \\
$\EAREO$ & $\Ex\sq{\frac{|U_{j}-V(q)_{j}| }{ U_{j} }}$  &   & Expected absolute relative error (to observation). \\
$\RMSRE_\tau$ & $\sqrt{\Ex\sq{\prn{\frac{|U_{j}-V(q)_{j}|}{\max(\tau, V(q)_{j})}}^2}}$ & $\tau \in \{3, 5,10\}$ & Root mean squared relative error at threshold $\tau$. \\
\bottomrule
\end{tabular}
\caption{\boldmath Error metrics considered for noise impact measurement. $V(q)_{j}$ is observed and $V(q)_{j}$ true value.}
\label{table:metrics_list}
\end{table*}

\newcommand{\greencheck}{\textcolor{green}{\checkmark}}
\newcommand{\orangecheck}{\textcolor{green}{\checkmark}}
\newcommand{\redx}{\textcolor{red}{x}}

\begin{table*}
\centering
\begin{tabular}{lccccccccc}
\toprule
Metric $\longrightarrow$ & $\ARE_{\alpha}$ & $\AME_\tau$ & $\APME_{\alpha, \tau}$ & $\EARE$ & $\RMSE$ & $\RMSRE$ & $\EARE_\tau$ & $\EAREO$ & $\RMSRE_\tau$ \\
\midrule
Decision Stability & \redx & \greencheck & \greencheck & \greencheck & \greencheck & \greencheck & \greencheck & \greencheck & \greencheck \\
Utility Stability & \orangecheck & \greencheck & \orangecheck & \orangecheck & \orangecheck & \greencheck & \greencheck & \redx & \greencheck \\
Ease of Optimization & \orangecheck & \orangecheck & \redx & \orangecheck & \greencheck & \greencheck & \orangecheck & \redx & \greencheck \\
Ease of Agg. Extension & \redx & \redx & \greencheck & \redx & \greencheck & \greencheck & \redx & \redx & \greencheck \\
Defined at Zero & \redx & \redx & \greencheck & \redx & \redx & \greencheck & \greencheck & \greencheck & \greencheck \\
Differentiates Small / Large & \greencheck & \redx & \greencheck & \greencheck & \greencheck & \redx & \greencheck & \greencheck & \greencheck \\
\bottomrule
\end{tabular}
\caption{Desirable properties for different metrics.}
\label{table:metrics_features}
\end{table*}

\newpage
\appendix

\section{Error metrics for evaluating reports} \label{sec:error-metrics}

In \Cref{table:metrics_list} above, we specify all metrics considered for utility evaluation.

To choose a particular metric,
we considered the desirable properties of an error metric that further can be used as an objective function. Ideally, a good error metric should have the following properties:
\begin{enumerate}[leftmargin=*]
\item\textbf{Decision Stability}: 
Some of our metrics are parameterized. (E.g., $\RMSRE_{\btau}$ is parameterized by $\btau$; see \Cref{def:rmse-tau}.) For a good metric, the decision from our optimization procedure (e.g, count bound, contribution budgeting, etc.) should not be too sensitive to the choice of parameters. 
\item\textbf{Utility Stability}: The utility measured by the metric is robust to perturbations in the input (e.g., true conversion count). For instance, the metric’s output should not change too much if the true conversion count is slightly changed. This is important because the true conversion count is often difficult to measure accurately.
\item\textbf{Ease of Optimization}: The metric is easy to calculate, and the objective function based on it is easy to be optimized.
\item\textbf{Ease of Extension}:  The metric should be easy to extend to aggregates after keyspace aggregation. The metric is for a slice in aggregate API, which is any possible combination of keys (keyspace value). For example, a slice for an advertiser may look like: \texttt{impression\_date=`8/1', biddability=`True'}. To get the total number of biddable conversions, one needs to sum up all noised counts from slices with \texttt{biddability=`True'}. This accumulates a bunch of Laplace noise random variables, which is no longer Laplace. It is desirable that the slice error metric can be easily adapted to aggregates after keyspace aggregation. \pasin{Does this ``ease of extension make sense?''}
\item\textbf{Defined at Zero}: The metric should be well-defined when conversion query value is zero. This is important since the conversion data can be sparse. \label{item:def-at-zero} 
\item\textbf{Differentiates Small/Large Values}: The metric should differentiate between large and small query values. Intuitively, this is because noise added to large values will usually have less effect on downstream tasks compared to the same amount of noise added to small values. \label{item:diff-small-large-val}   
\pritish{If we are not listing {\em Workable} as a point, then it must be removed from the table as well?}
\hidayet{I were not sure whether to include or not. as all metrics satisfied it. but I am going to include it as this was a relevant point to consider and readers might benefit from it.}
\end{enumerate}

\Cref{table:metrics_features} provides the list of criteria that each metric satisfies.

\paragraph{Intuition for $\RMSRE_{\tau}$.} $\RMSRE_{\tau}$ can be seen as a hybrid between additive and multiplicative error. When the query values are smaller than the threshold $\tau$, it becomes (a scaled version of) the root mean squared error $\RMSE$. Recall that 
\begin{align*}
\RMSE(u, q; D) &\displaystyle~:=~ \sqrt{\frac{1}{m} \sum_{j \in [m]} \Ex \prn{u_{j} - V_D(q)_{j}}^2},
\end{align*}
Meanwhile, if the query values are larger than $\tau$, then it becomes the root mean square relative error, defined as
\begin{align*}
\RMSRE(u, q; D) &\displaystyle~:=~ \sqrt{\frac{1}{m} \sum_{j \in [m]} \Ex \prn{\frac{u_{j} - V_D(q)_{j}}{V_D(q)_{j}}}^2},
\end{align*}
To give an intuition as to why $\RMSRE_{\tau}$ is a good metric, we can compare them with $\RMSE$ and $\RMSRE$. The main advantage of $\RMSRE_{\tau}$ over $\RMSE$ is that $\RMSRE_{\tau}$ can distinguish between the small and large values (criteria (\ref{item:diff-small-large-val}) above). Meanwhile, $\RMSRE_{\tau}$ is defined even when the query values are zero, whereas $\RMSRE$ is undefined (criteria (\ref{item:def-at-zero}) above).

\end{document}

%% file: macros.tex
\usepackage{amsmath,amsthm,amsfonts,bm}  
\usepackage{graphicx} 
\usepackage{enumitem}
\usepackage{multirow}
\usepackage{algorithm}
\usepackage[noend]{algorithmic}
\usepackage{xspace}
\usepackage{booktabs}  
\usepackage{wrapfig}  
\usepackage{xcolor}  
\usepackage{subfigure}

\usepackage{hyperref}
\hypersetup{
    colorlinks=true,
    linkcolor=blue,
    citecolor=blue,
    filecolor=blue,
    urlcolor=blue
}
\usepackage[nameinlink]{cleveref}
\usepackage{tikz}
\usetikzlibrary{calc,arrows.meta}

\setlist[itemize]{label=$\triangleright$}

\providecommand{\Comments}{0}  
\ifnum\Comments=1
\usepackage[colorinlistoftodos,prependcaption,textsize=scriptsize]{todonotes}
\paperwidth=\dimexpr \paperwidth + 5.6cm\relax
\oddsidemargin=\dimexpr\oddsidemargin + 2.8cm\relax
\evensidemargin=\dimexpr\evensidemargin + 2.8cm\relax
\marginparwidth=\dimexpr\marginparwidth + 3.4cm\relax
\else
\usepackage[disable]{todonotes}
\fi
\newcommand{\mytodo}[1]{\ifnum\Comments=1{#1}\fi}

\newcommand{\hidayet}[1]{\todo[linecolor=myPurple,backgroundcolor=myPurple!25,bordercolor=myPurple]{Hidayet: #1}}
\newcommand{\pritish}[1]{\todo[linecolor=myGold,backgroundcolor=myGold!25,bordercolor=myGold]{Pritish: #1}}
\newcommand{\pritishinfo}[1]{\todo[linecolor=Ggreen,backgroundcolor=Ggreen!25,bordercolor=Ggreen]{Pritish: #1}}

\newcommand{\pasin}[1]{\todo[linecolor=Gblue,backgroundcolor=Gblue!40,bordercolor=Gblue]{Pasin: #1}}

\newcommand{\tableoftodos}{\ifnum\Comments=1 \listoftodos[Comments/To Do's] \fi}

\definecolor{Gred}{RGB}{219, 50, 54}
\definecolor{Ggreen}{RGB}{60, 186, 84}
\definecolor{Gblue}{RGB}{72, 133, 237}
\definecolor{Gyellow}{RGB}{247, 178, 16}
\definecolor{ToCgreen}{RGB}{0, 128, 0}
\definecolor{myGold}{RGB}{231,141,20}
\definecolor{myBlue}{rgb}{0.19,0.41,.65}
\definecolor{myPurple}{RGB}{175,0,124}

\newtheorem{theorem}{Theorem}[section]
\newtheorem{observation}[theorem]{Observation}
\newtheorem{lemma}[theorem]{Lemma}
\newtheorem{relaxation}[theorem]{Relaxation}
\theoremstyle{definition}
\newtheorem{definition}[theorem]{Definition}

\newcommand{\prn}[1]{\left ( #1 \right )}
\newcommand{\sq}[1]{\left [ #1 \right ]}
\newcommand{\ceil}[1]{\left \lceil #1 \right \rceil}
\newcommand{\floor}[1]{\left \lfloor #1 \right \rfloor}
\newcommand{\pabs}[1]{\left| #1 \right|}

\DeclareMathOperator{\Ex}{\mathbb{E}}
\DeclareMathOperator{\Var}{\mathsf{Var}}
\DeclareMathOperator{\argmin}{\mathrm{argmin}}

\newcommand{\btau}{{\bm \tau}}

\newcommand{\bu}{{\bm u}}
\newcommand{\bv}{{\bm v}}

\newcommand{\eps}{\varepsilon}

\newcommand{\R}{\mathbb{R}}
\newcommand{\Z}{\mathbb{Z}}

\newcommand{\cA}{\mathcal{A}}
\newcommand{\cC}{\mathcal{C}}
\newcommand{\cD}{\mathcal{D}}
\newcommand{\cR}{\mathcal{R}}
\newcommand{\cS}{\mathcal{S}}
\newcommand{\cW}{\mathcal{W}}
\newcommand{\cX}{\mathcal{X}}
\newcommand{\cY}{\mathcal{Y}}
\newcommand{\cZ}{\mathcal{Z}}

\newcommand{\ow}{\overline{w}}
\newcommand{\ocW}{\overline{\mathcal{W}}}
\newcommand{\ocZ}{\overline{\mathcal{Z}}}

\newcommand{\RMSRE}{\mathsf{RMSRE}}
\newcommand{\CRR}{\mathsf{CRR}}
\newcommand{\RR}{\mathsf{RR}}
\DeclareMathOperator{\DLap}{\mathrm{DLap}}

\newcommand{\contribbudget}{\Gamma}

\newcommand{\clip}{\mathrm{clip}}

\newcommand{\E}{\mathbb{E}}
\newcommand{\cDc}{\cD_{\mathrm{count}}^{\mathrm{imp}}}
\newcommand{\cDcc}{\cD_{\mathrm{count}}^{\mathrm{conv}}}
\newcommand{\cDv}{\cD_{\mathrm{value}}^{\mathrm{conv}}}

\newcommand{\rem}{\mathrm{rem}}
\newcommand{\bias}{\mathrm{bi}}
\newcommand{\numabove}{\mathrm{nc}}
\newcommand{\ind}{\mathbf{1}}

\newcommand{\tC}{\tilde{C}}

\newcommand{\tR}{\tilde{R}}
\newcommand{\N}{\mathbb{N}}
\newcommand{\cDcom}{\mathcal{D}_{\mathrm{comp}}}

\newcommand{\ContributionCapping}{\textsf{ContributionBounding}\xspace}
\newcommand{\SummaryReport}{\textsf{SummaryReport}\xspace}
\newcommand{\HistogramContribution}{\textsf{HistogramContribution}\xspace}
\newcommand{\ReconstructValues}{\textsf{ReconstructValues}\xspace}
\newcommand{\ParameterOptimization}{\textsf{ParameterOptimization}\xspace}

\newcommand{\CSSCL}{CSSCL}  

%% file: main.bbl
\begin{thebibliography}{10}

\bibitem{aggregation-service-tee}
Aggregation service for the attribution reporting api.
\newblock
  \url{https://github.com/WICG/attribution-reporting-api/blob/main/AGGREGATION_SERVICE_TEE.md}.

\bibitem{aggregate-api-android}
Attribution reporting: Aggregatable reports api.
\newblock
  \url{https://developer.android.com/design-for-safety/privacy-sandbox/attribution#aggregatable-reports-api}.

\bibitem{privacy-guardrails-aggregate-api}
{Attribution Reporting API with Aggregatable Reports: Privacy Considerations}.
\newblock \url{
  https://github.com/WICG/attribution-reporting-api/blob/main/AGGREGATE.md#privacy-considerations}.

\bibitem{event-api-android}
{Attribution Reporting: Event-level Reports}.
\newblock \url{
  https://developer.android.com/design-for-safety/privacy-sandbox/attribution#event-level-reports}.

\bibitem{contribution-budget-ara}
Contribution budget for summary reports.
\newblock
  \url{https://developer.chrome.com/docs/privacy-sandbox/attribution-reporting/contribution-budget/}.

\bibitem{MaskedLARk-github}
{MaskedLARk}.
\newblock \url{https://github.com/microsoft/maskedlark}.

\bibitem{scipyopt}
scipy.optimize.minimize.

\bibitem{SKAdNetwork}
{SKAdNetwork}.
\newblock
  \url{https://developer.apple.com/documentation/storekit/skadnetwork/}.

\bibitem{amin2019bounding}
{\sc Amin, K., Kulesza, A., Munoz, A., and Vassilvtiskii, S.}
\newblock Bounding user contributions: A bias-variance trade-off in
  differential privacy.
\newblock In {\em Proceedings of the 36th International Conference on Machine
  Learning\/} (09--15 Jun 2019), K.~Chaudhuri and R.~Salakhutdinov, Eds.,
  vol.~97 of {\em Proceedings of Machine Learning Research}, PMLR,
  pp.~263--271.

\bibitem{clauset2009power}
{\sc Clauset, A., Shalizi, C.~R., and Newman, M.~E.}
\newblock Power-law distributions in empirical data.
\newblock {\em SIAM review 51}, 4 (2009), 661--703.

\bibitem{cormode2012differentially}
{\sc Cormode, G., Procopiuc, C., Srivastava, D., Shen, E., and Yu, T.}
\newblock Differentially private spatial decompositions.
\newblock In {\em ICDE\/} (2012), pp.~20--31.

\bibitem{dawson2023optimizing}
{\sc Dawson, M., Ghazi, B., Kamath, P., Kumar, K., Kumar, R., Luan, B.,
  Manurangsi, P., Mundru, N., Nair, H., Sealfon, A., and Zhu, S.}
\newblock {Optimizing Hierarchical Queries for the Attribution Reporting API}.
\newblock In {\em AdKDD\/} (2023).

\bibitem{DworkMNS06}
{\sc Dwork, C., McSherry, F., Nissim, K., and Smith, A.~D.}
\newblock Calibrating noise to sensitivity in private data analysis.
\newblock In {\em TCC\/} (2006), pp.~265--284.

\bibitem{dwork2014algorithmic}
{\sc Dwork, C., and Roth, A.}
\newblock The algorithmic foundations of differential privacy.
\newblock {\em Foundations and Trends{\textregistered} in Theoretical Computer
  Science 9}, 3--4 (2014), 211--407.

\bibitem{hay2009boosting}
{\sc Hay, M., Rastogi, V., Miklau, G., and Suciu, D.}
\newblock Boosting the accuracy of differentially-private histograms through
  consistency.
\newblock {\em VLDB\/} (2010).

\bibitem{kamath2023biasvarianceprivacy}
{\sc Kamath, G., Mouzakis, A., Regehr, M., Singhal, V., Steinke, T., and
  Ullman, J.}
\newblock A bias-variance-privacy trilemma for statistical estimation, 2023.

\bibitem{noise-lab}
{\sc Nadan, A., White, A., Cucu, A., Nalpas, M., and Mastromatto, Z.}
\newblock {Experiment with summary report design decisions}, November 2022.
\newblock
  \url{https://developer.chrome.com/docs/privacy-sandbox/summary-reports/design-decisions/}.

\bibitem{chrome-attribution-reporting}
{\sc Nalpas, M., and White, A.}
\newblock {Attribution Reporting}, May 2021.
\newblock
  \url{https://developer.chrome.com/en/docs/privacy-sandbox/attribution-reporting/}.

\bibitem{pfeiffer2021masked}
{\sc Pfeiffer~III, J.~J., Charles, D., Gilton, D., Jung, Y.~H., Parsana, M.,
  and Anderson, E.}
\newblock Masked lark: Masked learning, aggregation and reporting workflow.
\newblock {\em arXiv preprint arXiv:2110.14794\/} (2021).

\bibitem{chromium}
{\sc Schuh, J.}
\newblock {Building a more private web: A path towards making third party
  cookies obsolete}, January 2020.
\newblock
  \url{https://blog.chromium.org/2020/01/building-more-private-web-path-towards.html}.

\bibitem{tallis2018reacting}
{\sc Tallis, M., and Yadav, P.}
\newblock Reacting to variations in product demand: An application for
  conversion rate {(CR)} prediction in sponsored search.
\newblock In {\em IEEE BigData\/} (2018).

\bibitem{ipa-blog}
{\sc Thomson, M.}
\newblock {Privacy Preserving Attribution for Advertising}, February 2022.
\newblock
  \url{https://blog.mozilla.org/en/mozilla/privacy-preserving-attribution-for-advertising/}.

\bibitem{safari}
{\sc Wilander, J.}
\newblock {Full Third-Party Cookie Blocking and More}, March 2020.
\newblock
  \url{https://webkit.org/blog/10218/full-third-party-cookie-blocking-and-more/}.

\bibitem{pcm-safari}
{\sc Wilander, J.}
\newblock {Introducing Private Click Measurement, PCM}, February 2021.
\newblock
  \url{https://webkit.org/blog/11529/introducing-private-click-measurement-pcm/}.

\bibitem{mozilla}
{\sc Wood, M.}
\newblock {Today’s Firefox Blocks Third-Party Tracking Cookies and
  Cryptomining by Default}, 2019.
\newblock
  \url{https://blog.mozilla.org/en/products/firefox/todays-firefox-blocks-third-party-tracking-cookies-and-cryptomining-by-default/}.

\end{thebibliography}
